\documentclass[fleqn,usenatbib]{mnras}

\usepackage{newtxtext,newtxmath}

\usepackage[T1]{fontenc}
\usepackage[utf8]{inputenc}

\DeclareRobustCommand{\VAN}[3]{#2}
\let\VANthebibliography\thebibliography
\def\thebibliography{\DeclareRobustCommand{\VAN}[3]{##3}\VANthebibliography}

\usepackage{graphicx}	
\usepackage{amsmath}	
\usepackage{physics}    
\usepackage{ulem}       
\usepackage{bigints}
\usepackage{fix-cm}

\newcommand{\vect}[1]{\boldsymbol{#1}}

\newcommand{\se}[1]{Section~\ref{sec:#1}}
\newcommand{\ses}[1]{Sections~\ref{sec:#1}}

\newcommand{\fig}[1]{Fig.~\ref{fig:#1}}
\newcommand{\figs}[1]{Figs.~\ref{fig:#1}}
\newcommand{\figss}[1]{\ref{fig:#1}}
\newcommand{\Fig}[1]{Figure~\ref{fig:#1}}

\newcommand{\tab}[1]{Table~\ref{tab:#1}}
\newcommand{\be}{\begin{equation}}
\newcommand{\ee}{\end{equation}}
\newcommand{\bea}{\begin{eqnarray}}
\newcommand{\eea}{\end{eqnarray}}

\newcommand{\no}{\noindent}

\newcommand{\msun}{{\rm M}_\odot}
\newcommand{\Msun}{M_\odot}

\newcommand{\ifm}[1]{\relax\ifmmode#1\else$\mathsurround=0pt #1$\fi}
\newcommand{\kms}{\ifmmode\,{\rm km}\,{\rm s}^{-1}\else km$\,$s$^{-1}$\fi}

\newcommand{\Mpc}{\,{\rm Mpc}}
\newcommand{\kpc}{\,{\rm kpc}}
\newcommand{\pc}{\,{\rm pc}}

\newcommand{\yr}{\,{\rm yr}}

\newcommand{\Myr}{\,{\rm Myr}}

\newcommand{\ltsima}{$\; \buildrel < \over \sim \;$}
\newcommand{\lsim}{\lower.5ex\hbox{\ltsima}}
\newcommand{\gtsima}{$\; \buildrel > \over \sim \;$}
\newcommand{\gsim}{\lower.5ex\hbox{\gtsima}}

\def\la{\langle}

\def\cmc{\,{\rm cm}^{-3}}
\def\cms{\,{\rm cm}^{-2}}

\def\M*{M_{\rm *}}

\def\Kdegree{{\rm K}}

\def\Pi{\varpi_{_{\rm I}}}

\usepackage{pifont}

\usepackage{scalerel}
\newcommand\HI{\text{H}\protect\scaleto{$I$}{1.2ex}}

\newcommand\OVI{\text{O}\protect\scaleto{$VI$}{1.2ex}}

\title[CGM and CR Models]{Constraining cosmic ray transport models using circumgalactic medium properties and observables}

\author[Y. S. Lu et al.]{\parbox[t]{\textwidth}{Yue Samuel Lu,$^{1,2}$\thanks{E-mail: yul232@ucsd.edu}
Dušan Kereš,$^{1, 2}$
Philip F. Hopkins,$^{3}$
Sam B. Ponnada, $^{3}$
Claude-André Faucher-Giguère,$^{4}$
Cameron B. Hummels$^{3}$ }
\vspace{0.2cm}\\%
$^{1}$Department of Astronomy and Astrophysics, University of California, San Diego, La Jolla, CA 92093, USA\\%
$^{2}$Department of Physics, University of California, San Diego, La Jolla, CA 92093, USA\\%
$^{3}$TAPIR, California Institute of Technology, Mailcode 350-17, Pasadena, CA 91125, USA\\%
$^{4}$Department of Physics and Astronomy and CIERA, Northwestern University, 2145 Sheridan Road, Evanston, IL 60208, USA\\%
}

\date{Accepted XXX. Received YYY; in original form ZZZ}

\pubyear{2024}

\begin{document}
\label{firstpage}
\pagerange{\pageref{firstpage}--\pageref{lastpage}}
\maketitle

\begin{abstract}

Cosmic rays (CRs) are a pivotal non-thermal component of galaxy formation and evolution. However, the intricacies of CR physics, particularly how they propagate in the circumgalactic medium (CGM), remain largely unconstrained. In this work, we study CGM properties in FIRE-2 (Feedback In Realistic Environments) simulations of the same Milky Way (MW)-mass halo at $z=0$ with different CR transport models that produce similar diffuse $\sim$ GeV $\gamma$-ray emission, as an attempt to further constrain CR transport models. We study the gas morphology and thermal properties, and generate synthetic observations of rest-frame UV ion absorption columns and X-ray emission. CRs lower galaxy masses and star formation rates (SFRs) while supporting more cool CGM gas, which boosts the \text{H}\protect\scaleto{$I$}{1.2ex} and \text{O}\protect\scaleto{$VI$}{1.2ex} column densities in the CGM, bringing simulations more in line with observations, but there can be large differences between CR transport models and resolution levels. X-ray emission within and close to galaxies is consistent with thermal (free-free and metal-line) emission plus X-ray binaries, while more extended ($\sim 100\,$kpc) CGM emission is potentially dominated by inverse Compton scattering (ICS), motivating future work on the spatially resolved X-ray profiles. Although comparisons with observations are sensitive to sample selection and mimicking the details of observations, and our analysis did not result in strong constraints on CR models, the differences between simulations are significant and could be used as a framework for future studies.

\end{abstract}




\section{Introduction}
\label{sec:intro}

The circumgalactic medium (CGM), the gas that fills the space between a galaxy and the boundary of its host halo, is one of the most important and yet complicated parts in building our understanding of galaxy evolution. It serves as the main reservoir for gas accreted onto the galaxy to form stars and for gas recycled back into the CGM from various feedback processes \citep{tumlinson2017circumgalactic,faucher2023key}. One particularly interesting feedback mechanism is the cosmic rays (CRs), that can provide non-thermal support to the CGM gas and regulate its phase structure. The interaction between CRs and the CGM gas involves numerous physical processes across different scales \citep{guo2008feedback, butsky2018role, 2014ApJ...797L..18S, hopkins2020but, ji2020properties, ruszkowski2023cosmic} that are not yet fully understood.

Many efforts have been made to detect and understand the composition and kinematics of the CGM, both using observational and theoretical tools. For Milky Way (MW)-mass galaxies, the early models assumed that ``hot'' ($\gsim 10^6 \Kdegree$) gas is the main component of the CGM, filling the dark matter (DM) haloes \citep{spitzer1953transport, white1978core}, whose existence was supported by X-ray observations (e.g., \citet{li2013chandra}, \citet{fang2012hot}) and $\OVI$ absorption features (e.g., \citet{wang2005warm}). With the help of modern supercomputer simulations, it is now widely believed that ``cold'' ($\sim 10^4 \Kdegree$) filamentary streams and cold mode accretions at high redshifts are the major modes feeding MW-mass galaxies \citep{kerevs2005galaxies, dekel2009cold}, but at low redshifts, cold filaments are expected to be heated up in haloes of $L_*$ galaxies. Therefore, the most timely picture of the CGM gas should describe it as a combination of cold, warm, and hot gas, which are all detected in multiple observations \citep{stocke2013characterizing, werk2013cos, prochaska2017cos}. However, the distinctions of cold/cool, warm, and hot gas are not always well-defined and the temperature ranges associated with them can be somewhat nebulous. The CGM gas is not only multiphased and multiscaled, but also exhibits complicated kinematics, undergoing instabilities (e.g., cooling and rotational instabilities; see \citet{mccourt2018characteristic,stern2019cooling,lochhaas2020properties}) and exhibiting strong inflow/outflow features \citep{steidel2010structure, muratov2017metal, trapp2022gas}. Furthermore, the gas is supported by various forces (both thermal and non-thermal) against its own self-gravity, leading to significant impacts on the CGM phase structure \citep{hopkins2020but, ji2020properties, zhang2023inspiraling}.

Feedback processes and fluid microphysics are critical in galaxy evolution models, including stellar and supernova feedback \citep{larson1974effects, angles2017black}, black hole and AGN feedback \citep{hopkins2005black, hopkins2008cosmological, faucher2022agn, byrne2023formation}, viscosity and conduction \citep{reynolds2005buoyant, hopkins2020but}, magnetic fields (B-fields) \citep{beck2016magnetic}, dust \citep{choban2022galactic}, and the focus of this work, cosmic rays \citep{fujita2011stable, wiener2013cosmic, pfrommer2017simulating}. There are now many approaches to implement feedback processes in galaxy evolution simulations, among which one of the most successful implementations is the Feedback In Realistic Environments (FIRE) project\footnote{\hyperlink{https://fire.northwestern.edu}{https://fire.northwestern.edu}} \citep{hopkins2014galaxies, hopkins2018fire, hopkins2023fire}. In this work, we use the simulated galaxies from the FIRE-2 version. We summarize the simulation methods in \se{Methods}.

Due to a lack of observational constraints on the detailed properties of B-fields on small ($\ll{\rm pc}$) scales, the transport of CRs through the ISM and CGM remains poorly constrained. This is why previous work often assumed a constant (both spatially and temporally) isotropic diffusivity $\kappa_{\rm iso}$, as presented in \citet{chan2019cosmic, hopkins2021cosmic}. Furthermore, given that the CR energy density is dominated by the $\sim 1-10 {\rm \, GeV}$ protons, most previous works adapted a single energy bin for CRs and transport properties relevant for those energies. However, it is clear that constant diffusivity lacks a proper physical motivation \citep{hopkins2021effects, hopkins2021testing}. Motivated by the work on small-scale CR transport, \citet{hopkins2021testing} has recently implemented several different physical models for CR transport on resolved scales in FIRE-2 simulations. In addition, full-spectrum CR physics (as opposed to the single energy bin) was also implemented and coupled to the FIRE model in \citet{hopkins2022first}. Different groups, using different codes, have also implemented CR transport physics to varying degrees of complexity (e.g., \texttt{AREPO}, see \citet{thomas2023cosmic} and \citet{buck2020effects}; \texttt{ENZO}, see \citet{butsky2018role}).

The details of the physically motivated CR transport models from \citet{hopkins2021testing} were thoroughly studied in previous works (e.g., \citet{jokipii1966cosmic, skilling1975cosmic}). However, because of the uncertainties of magnetic fields and CR microphysics on ($\sim \text{au}$) scales, FIRE-2 can only implement such models macroscopically, scaling the diffusivity $\kappa$ by macroscopic plasma properties (see \se{2.2} for more details). This leaves an open question: which of the physically motivated models can give rise to the correct emergent large-scale transport of CRs? Previous studies suggested that observables sensitive to the CGM phase structure could be good tracers of the influence of CRs on the surrounding gas. In \citet{ji2020properties}, a thorough analysis of the CGM was performed with regard to its thermal properties and phase structures, focusing on the distinction between CR and non-CR runs. In \citet{butsky2023constraining}, the effective diffusivity in the CGM $\kappa_{\rm eff}$ was constrained using the total $\HI$ column inferred by optical/UV absorption line spectroscopy from the COS-Halos survey in tandem with a simple analytic model for the CR pressure in the CGM. In \citet{hopkins2021effects}, several preliminary CR transport models with variant diffusivity were selected based on basic order-of-magnitude comparisons, and those models were used in several subsequent works, including \citet{ponnada2023synchrotron, ponnada2024synchrotron}, in which the authors studied synchrotron emission from galaxies (see \se{2.3} for more detailed description of those runs). In this paper, we plan to extend these early works and use properties of CGM gas to constrain the CR transport models.

In this paper, our aim is to systematically constrain CR models using physical properties and observables of the CGM. The paper is organized as follows. In \se{Methods} we introduce our simulation methods, selected simulations, and the three major CR transport models that we will try to constrain. In \se{Basic Prop} we present the basic properties of the galaxy of our selected simulations, including their gas morphological structure and thermal profiles. In \se{Absorption} we post-process our simulations to reproduce rest-frame UV ion absorption columns of CGM and X-ray emission luminosities and then compare the results from our simulations to observations. In \se{Conclusions} we conclude our findings and discuss future work. 


\section{Methods}
\label{sec:Methods}
In this section, we describe the key physics in FIRE-2 simulations (\se{2.1}) and CR transport (\se{2.2}) with a particular focus on the three major transport models: constant diffusivity (CD), extrinsic turbulence (ET), and self-confinement (SC). In \se{2.3}, we present the simulations we selected for this analysis.

\subsection{FIRE-2 Simulations}
\label{sec:2.1}
We use simulations evolved with the FIRE-2 method for star formation and feedback \citep{hopkins2018fire}, which is an updated version of the FIRE model \citep{hopkins2014galaxies}. Here, we briefly describe the FIRE-2 methods and refer the readers to the original references for further details. The simulations are run with the \texttt{GIZMO} code \citep{hopkins2015new}, with a meshless finite-mass MFM solver (a mesh-free finite-volume Lagrangian Godunov method). All simulations include ideal magnetohydrodynamics (MHD, except for Hydro+ runs, see \se{2.3}), thermal conduction and viscosity \citep{hopkins2020but}, and gravity with adaptive Lagrangian force softening for gas. These are cosmological ``zoom-in'' simulations, evolving a large box from $z \gsim 100$ with resolution concentrated on a $\sim 1 - 10 \Mpc$ comoving volume around a target halo of interest.

Radiative cooling, star formation, and stellar feedback mechanisms are all included following the FIRE-2 implementation \citep{hopkins2018fire}. Gas cooling is followed from $10-10^{10} \Kdegree$ including Compton, free-free, metal line, fine-structure, atomic, and molecular processes. We also follow photoelectric and photoionization gas heating by both local sources
and an uniform metagalactic background including the effect of self-shielding \citep{faucher2009new}. As for stars, star formation is allowed when all the following criteria are satisfied: the gas is locally self-gravitating \citep{hopkins2013meaning, grudic2018feedback}, self-shielding, Jeans-unstable, molecular, and dense ($n >1000 \cmc$). Stars then evolve according to standard star evolution tracks, assuming Kroupa IMF \citep{kroupa2011stellar} and explicitly accounting for the mass, metals, momentum, and energy injected via individual supernovae (SNe, type Ia \& II), O/B and AGB-star mass-loss \citep{hopkins2018model}, and their radiation \citep{hopkins2020radiative}.

We adopt a standard, flat $\Lambda$CDM cosmology with $h\approx 0.70$, $\Omega_{\rm M}=1-\Omega_{\Lambda}=0.27$, and $\Omega_{\rm b}\approx 0.045$, consistent with the current constraint \citep{2014A&A...571A..16P}.

\subsection{CR Physics}
\label{sec:2.2}
The complete prescription of CR physics implemented in our simulations can be found in \citet{chan2019cosmic}, \citet{hopkins2022consistent}, \citet{hopkins2020but}, and \citet{hopkins2021testing}. CRs are injected from SNe and OB/WR stellar winds with an energy efficiency of $\varepsilon_{\rm CR} = 0.1$ of the initial ejecta kinetic energy. Note that here we only evolve CRs with a ``single-bin'' ($\sim {\rm GeV}$) energy spectrum (or equivalently, a constant spectral distribution of CRs). We treat the CRs as a relativistic fluid (e.g., \citet{mckenzie1982non}) with energy density $e_{\rm cr}$ and pressure $P_{\rm cr}=e_{\rm cr}/3$ that obeys the two-moment transport equations:

\be
\pdv{e_{\rm cr}}{t}+\nabla\cdot\left[\vect{u}(e_{\rm cr} + P_{\rm cr})+\vect{F} \right]=\vect{u}\cdot \nabla P_{\rm cr}-\Lambda_{\rm st}-\Lambda_{\rm cool}+S_{\rm in},
\ee

\be
\frac{\mathbb{D}_t\vect{F}}{\tilde{c}^2}+\nabla_\parallel P_{\rm cr}=-\frac{\vect{F}}{3\kappa_*},
\ee

\no where $\vect{u}$ is the fluid velocity, $\vect{F}$ is the CR flux in the fluid frame, $S_{\rm in}$ is the CR source term (assuming $10\%$ of the SNe mechanical energy goes to CRs), $\nabla_\parallel = \vect{\hat{b}}( \vect{\hat{b}}\cdot \nabla)$ is the component of the gradient operator parallel to the magnetic field $\vect{B}$ (neglecting perpendicular diffusion), $\Lambda_{\rm st}=\min(v_{\rm A}, v_{\rm st})\left|\nabla_\parallel P_{\rm cr}\right|$ represents losses form streaming associated damping ($v_{\rm A}=\sqrt{B/4\pi \rho}$ is the Alfvén speed, and $v_{\rm st}$ is the streaming velocity), $\Lambda_{\rm cool}=5.8\times 10^{-16}\cmc {\rm \: s}^{-1}(n_{\rm n}+0.28n_{\rm e})e_{\rm cr}$ are the collisional losses (where $n_{\rm n}$ is nucleon number density for hardronic losses, and $n_{\rm e}$ is the free electron density for Coulomb losses \citep{guo2008feedback}), $\tilde{c}$ is the maximal CR free-streaming speed (similar to the reduced speed of light approximation in two-moment methods for radiation transport) $\mathbb{D}_t\vect{F}\equiv \vect{\hat{F}}\left[\partial |\vect{F}|/\partial t+\nabla\cdot (\vect{u}|\vect{F}|)+\vect{F}\cdot \left\{(\vect{\hat{F}} \cdot \nabla \vect{u})\right\}\right]$ is the derivative operator derived in \citet{thomas2019cosmic}, and $\kappa_*=\kappa_\parallel+(4/3)l_{\rm cr}v_{\rm st, \parallel}$ (where $l_{\rm cr}=P_{\rm cr}/|\nabla_\parallel P_{\rm cr}|$) contains the streaming and diffusion of CRs. On the other hand, CRs will influence the gas with $P_{\rm cr}$, and with a fraction of streaming+hadronic+Coulomb losses that are thermalized.

One can easily notice that the only uncertain part of the above prescription is the diffusivity $\kappa_*$\footnote{Other quantities either can be constrained from observations or arise from intrinsic properties of galaxies.}, which consists of two free parameters: $\kappa_\parallel$, the parallel diffusivity, and $v_{\rm st}$, the streaming velocity. A ``pure diffusion'' model of CR will have $\tilde{c}\rightarrow \infty$ \footnote{In practice, we set $\tilde{c}$ to be much greater than any possible velocity scale appearing in the simulation to reach this effect} and $v_{\rm st}=0$. Similarly, a ``pure streaming'' model will still have $\tilde{c}\rightarrow \infty$ but $\kappa_\parallel = 0$. Previous findings have concluded that, most parameters in our CR models are uncertain to a relatively modest level (factor $\sim 2$) but $\kappa_*$ can vary orders of magnitude without significantly impacting galaxy properties for the MW-mass galaxies alone. The exact details of the scaling of $\kappa_\parallel$ and $v_{\rm st}$ with local properties of gas and CRs can be complex, as shown and tested in \citet{hopkins2021testing}. From those results, several reasonable models were chosen and further tested for the basic galaxy properties in \citet{hopkins2021effects}. We choose our models based on their results to further use CGM to constrain the models. The models we consider in this work are presented below.

\begin{itemize}
    \item \textbf{Constant diffusivity (CD) model.} In this model, we set $\kappa_\parallel={\rm constant}$ and $v_{\rm st}=v_{\rm A}^{\rm ideal}$ (ideal MHD Alfvén wave).
    There is no clear physical motivation behind this model, but it is the most commonly adapted one. Our analysis includes the runs with $\kappa_\parallel =3\times 10^{29} {\rm \: cm}^2 {\rm \: s^{-1}}$ that provide a good match to galactic observables, such as diffuse $\gamma$-ray emission \citep{chan2019cosmic, hopkins2020but}.

    \item \textbf{Extrinsic turbulence (ET) model.} In this model, we assume CRs are scattered from ambient, preexisting turbulence \citep{jokipii1966cosmic}. The scaling of the model is $\kappa_\parallel \propto \mathcal{M}_{\rm A}^{-2}cl_{\rm turb}f_{\rm turb}$, where $\mathcal{M}_{\rm A}$ is the Alfvén Mach number, and $l_{\rm turb}$ is the turbulent length scale. In this paper, we present the models ``Alfvén-Max'' and ``Fast-Max''. For the ``Alfvén-Max'' model, we assume $f_{\rm turb}=1$, and for the ``Fast-Max'' model, $f_{\rm turb}\sim 30 \mathcal{M}_A^{5/3}{\rm Re}^{-1/3}$, where ${\rm Re}$ is the Reynolds number. The ``Fast-Max'' model describes the fast-mode scattering of CRs. The details of this model are described in Section 3.2 of \citet{hopkins2021testing}.
    
    \item \textbf{Self-confinement (SC) model.} In this model, we assume that the CRs are scattered from self-excited gyro-resonant Alfvén waves \citep{skilling1975cosmic}. We scale $\kappa_\parallel=cr_{\rm L}(16/3\pi)(l_{\rm turb}\Gamma_{\rm eff}/v_{\rm A}^{\rm ion})(e_{\rm B}/e_{\rm cr})f_{\rm QLT}$, where $r_{\rm L}$ is the Larmor radius, $e_{\rm B}=B^2/8\pi$ is the magnetic energy density, and $\Gamma_{\rm eff}$, which is the effective source term, contains different components that are essential for the scaling of the model. In this paper, we will consider three submodels: the $f_{\rm QLT}-100$ model, the $f_{\rm}-{\rm K41}$ model, and the $f_{\rm cas}-50$ model. Detailed scaling of the physics of these three models can be found in Section 3.3 of \citet{hopkins2021testing}.
\end{itemize}

Note that all of the aforementioned models (as well as all the variants that will be discussed later) satisfy galactic and local MW observational constraints (e.g., diffuse $\gamma$-ray emission), as their effective transports through the galaxy are all similar to the CR-CD model, but they can differ significantly in the CGM.

All of the aforementioned variant diffusivity CR models were also discussed in \citet{ponnada2023synchrotron}. The paper investigates the synchrotron emission from galaxies that results from underlying cosmic rays and magnetic fields, and it focuses on the variation between different CR models. We follow the simulations naming procedure therein \footnote{This is not completely consistent with the published namings in \citet{hopkins2021effects}. We therefore follow the naming from \citet{ponnada2023synchrotron} where any naming mismatches were corrected}.

Lastly, although all runs analyzed in this work evolve a single-bin CR energy spectrum, an analysis of full-spectrum CR runs using the implementation methods mentioned in \citet{hopkins2022first} will be left for future work.

\subsection{Selected Simulations}
\label{sec:2.3}

\begin{table*}
    \resizebox{\textwidth}{!}
        {\begin{tabular}{l l l l l l l l}
        \hline
        Target halo 	 & Basic physics 	 & Resolution 	 & $\mathrm{R}_{\mathrm{vir}}$ 	 &  $\mathrm{M}_{\mathrm{vir}}$	& $\mathrm{M}_{\star}(<0.1 R_{\rm vir})$ & $\kappa_{\rm CR}$ or  transport model 	 \\
             &  	 & \textit{high} or \textit{normal} 	 &  	[pkpc] & $[\mathrm{M}_{\odot}]$ 	 & $[\mathrm{M}_{\odot}]$ & [${\rm cm^2\: s^{-1}}$]	\\
        \hline
        \multicolumn{7}{c}{\textbf{Non-CR Runs}} \\
        \hline
        \texttt{m12i} 	 & FIRE-2 Hydro+ & high  & 269  & $1.1\times 10^{12}$  & $7.1\times 10^{10}$ & N/A     \\
        \texttt{m12i}     & FIRE-2 Hydro+    & normal    &  279  &  $1.2\times 10^{12}$ & $14.0\times 10^{10}$  & N/A    \\
        \texttt{m12i} 	 & FIRE-2 MHD+  & high   & 271  & $1.1\times 10^{12}$  &$8.0\times 10^{10}$ & N/A    \\
        \texttt{m12i}     & FIRE-2 MHD+    & normal   &  276  & $1.2\times 10^{12}$  & $13.3\times 10^{10}$  & N/A    \\

        \hline
        \multicolumn{7}{c}{\textbf{CR runs}} \\
        \hline
        
        \texttt{m12i}     & FIRE-2 CR+    & high    &  262  &  $1.0\times 10^{12}$ & $2.7\times 10^{10}$  & $3\times 10^{29}$         \\
        \texttt{m12i}     & FIRE-2 CR+   & normal    &  265  & $1.0\times 10^{12}$  & $5.9\times 10^{10}$ & $3\times 10^{29}$    \\
        
        \texttt{m12i}    & FIRE-2 CR+    & normal    &  264  & $1.0\times 10^{12}$  & $7.0\times 10^{10}$  & ET, Alfvén-Max    \\
        \texttt{m12i}    & FIRE-2 CR+    & normal    &  264  & $1.0\times 10^{12}$  & $6.7\times 10^{10}$  & ET, Fast-Max    \\
        \texttt{m12i}    & FIRE-2 CR+    & normal    &  272  &  $1.1\times 10^{12}$ &  $8.4\times 10^{10}$ & SC, $f_{\rm QLT}-100$  \\
        \texttt{m12i}    & FIRE-2 CR+    & normal    &  271  &  $1.1\times 10^{12}$ & $10.0\times 10^{10}$  & SC, $f_{\rm cas}-50$  \\
        \texttt{m12i}    & FIRE-2 CR+    & normal   &  273  &  $1.1\times 10^{12}$ & $11.0\times 10^{10}$  & SC, $f_{\rm cas}-{\rm K41}$   \\
        
        \hline
        \end{tabular}}
    \caption{The main FIRE-2 simulation runs analyzed in this work. The virial radii $R_{\rm vir}$ and the virial masses $M_{\rm vir}$ are calculated following \citet{bryan1998statistical}. 
    All properties shown here are for the $z=0$ snapshots. The average $R_{\rm vir}$ for our \texttt{m12i} haloes is $273 \kpc$ with an enclosed $M_{\rm vir}\approx 1.1\times 10^{12}\Msun$, close to the value for the standard target halo \texttt{m12i} \citet{hopkins2018fire}.}
    \label{tab:sims}
\end{table*}

We select a total of 11 simulation runs and divide them into two categories for our analysis: non-CR runs (including Hydro+ and MHD+) and CR runs (including CD, ET, and SC models). Each simulation run is of the halo \texttt{m12i} (we explore additional simulations with different target haloes in the Appendix \ref{apx:addtional}). Note that most runs have comparable resolution with most other simulations (in which each gas cell has $m_{\rm gas \, cell}\approx 56000 \Msun$, dubbed as \textit{``normal-resolution''} or \textit{``normal-res''} hereafter) but Hydro+, MHD+ and CR-CD runs are also available with FIRE-2 default high, ``latte'' resolution (in which each gas cell has $m_{\rm gas \, cell}\approx 7000 \Msun$, dubbed as \textit{``high-resolution''} or \textit{``high-res''} hereafter).\footnote{We recognize this naming may not be completely consistent with some previous FIRE papers (e.g., \citet{hopkins2018fire}, but since all our CR runs with variant diffusion models (ET and SC) are only available in the lower resolution, owing to high computing cost of higher resolution, this is the default resolution for our analysis and we refer to it as the "normal resolution".} We comment below in \se{3.1} that resolution effects are much less significant on CR runs compared to the Hydro+ and MHD+ runs. 

To locate the target halo of each simulation, we use GizmoAnalysis (\hyperlink{http://ascl.net/2002.015}{http://ascl.net/2002.015}), which was first used by \citet{wetzel2016reconciling}, to find the centres of the target haloes. Then we follow \citet{bryan1998statistical} to calculate the virial radius $R_{\rm vir}$ and the enclosed virial mass $M_{\rm vir}$ of each halo. \footnote{We note that all major halo information can be found in multiple FIRE papers, including the major FIRE-2 release \citet{hopkins2018fire}. For completeness purposes, we still perform our calculations without using any halo finder algorithms for our $z=0$ haloes, at which all galaxies have well-defined centres.} In \tab{sims} we categorize the simulations analyzed into the two categories mentioned above and show their $R_{\rm vir}$, $M_{\rm vir}$, and the central stellar mass $M_*(<0.1 R_{\rm vir})$, together with the key physics features of each run. Different physics models applied to the same target halo yield only slightly different $R_{\rm vir}$ and $M_{\rm vir}$. On the other hand, each galaxy's central stellar mass, $M_*(<0.1 R_{\rm vir})$, can vary significantly between variants. These signatures will be discussed in more detail later in \se{3.1} and can be used as a constraint for our CR models.


\section{Using Basic Galaxy Properties to Constrain CR Models}
\label{sec:Basic Prop}
In this section we present some basic galaxy properties at $z=0$, from which we conclude our preliminary constraints on the CR models. Here, we primarily focus on the difference between CR and non-CR runs. Due to the lack of statistics, we focus more on galaxy properties over which the CRs have more systematical effects. 

\subsection{Morphology of Galactic Gas}
\label{sec:3.1}

\begin{figure*}
    \centering
    \includegraphics[trim={0.0cm 0.0cm 0.0cm 0.0cm}, clip, width =0.98 \textwidth]{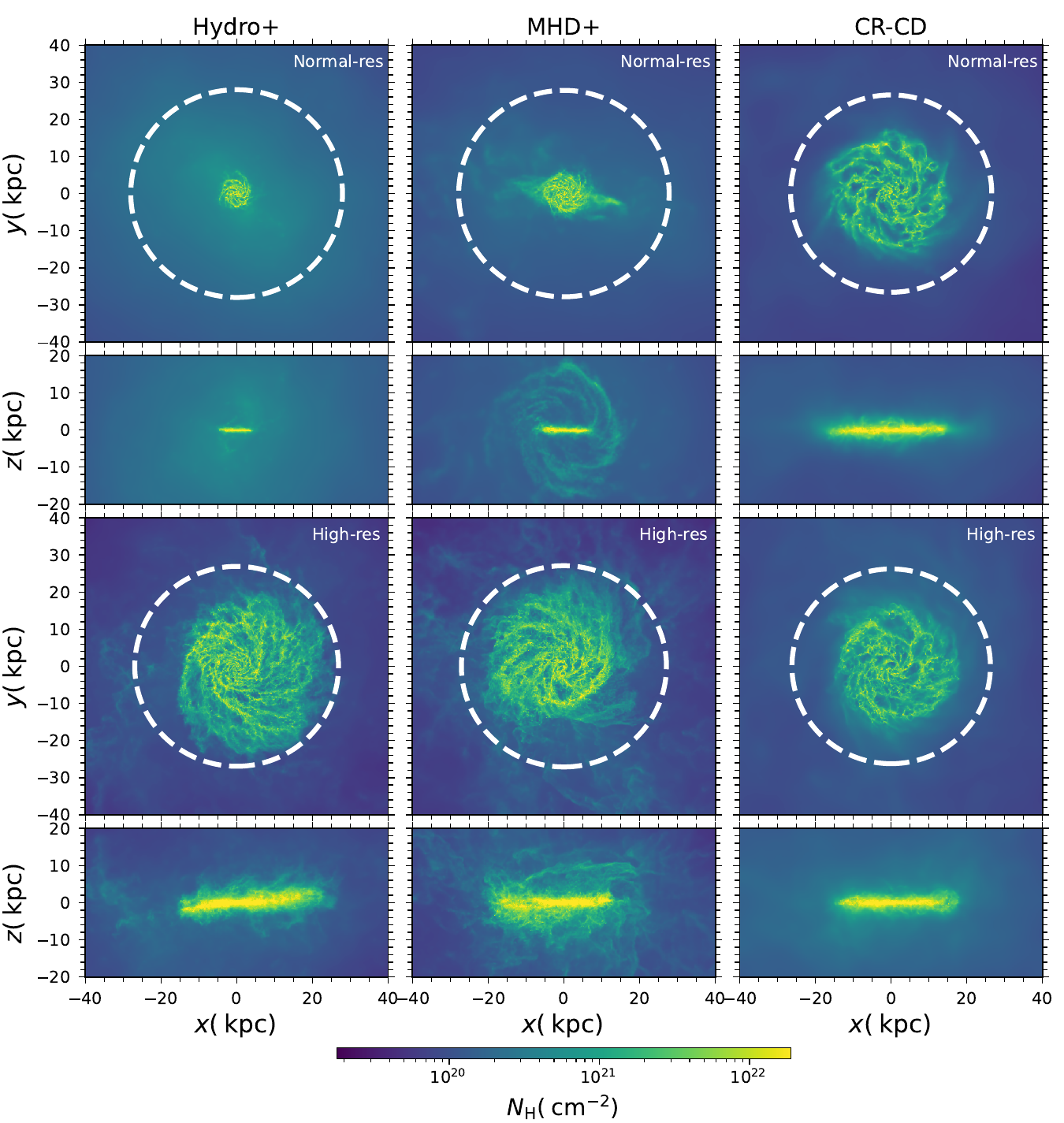}
    \caption{Maps of projected total hydrogen column density, $N_{\rm H}$, of our \texttt{m12i} simulation runs. The projections were taken over the whole simulation volumes. We include Hydro+, MHD+, and CR-CD runs. All maps are at $z=0$. We show the face-on (larger panels) and edge-on (smaller panels) projections. We compare the normal-res run ($m_{\rm gas \, cell}\approx 56000 \Msun$, upper half) with the high-res run ($m_{\rm gas \, cell}\approx 7000 \Msun$, lower half) of each simulation. The white dashed circles represent $0.1 R_{\rm vir}$ as listed in \tab{sims}. We see that all six runs form a galactic disk at $z=0$ to some extent, but while all three high-res runs show clear disks, only CR runs have the same structure of a symmetric disk at normal resolution. Furthermore, by comparing the upper half with the lower half, we see that the resolution has a much less significant effect on the CR-CD run than the Hydro+ and MHD+ runs. Normal-res Hydro+ and MHD+ runs locked almost all available halo baryons into stars (see \tab{sims}), resulting in a lack of late-time accretion to form a clear disk.}
\label{fig:proj_resolution_comparison}
\end{figure*}

\begin{figure*}
    \centering
    \includegraphics[trim={0.0cm 0.0cm 0.0cm 0.0cm}, clip, width =0.98 \textwidth]{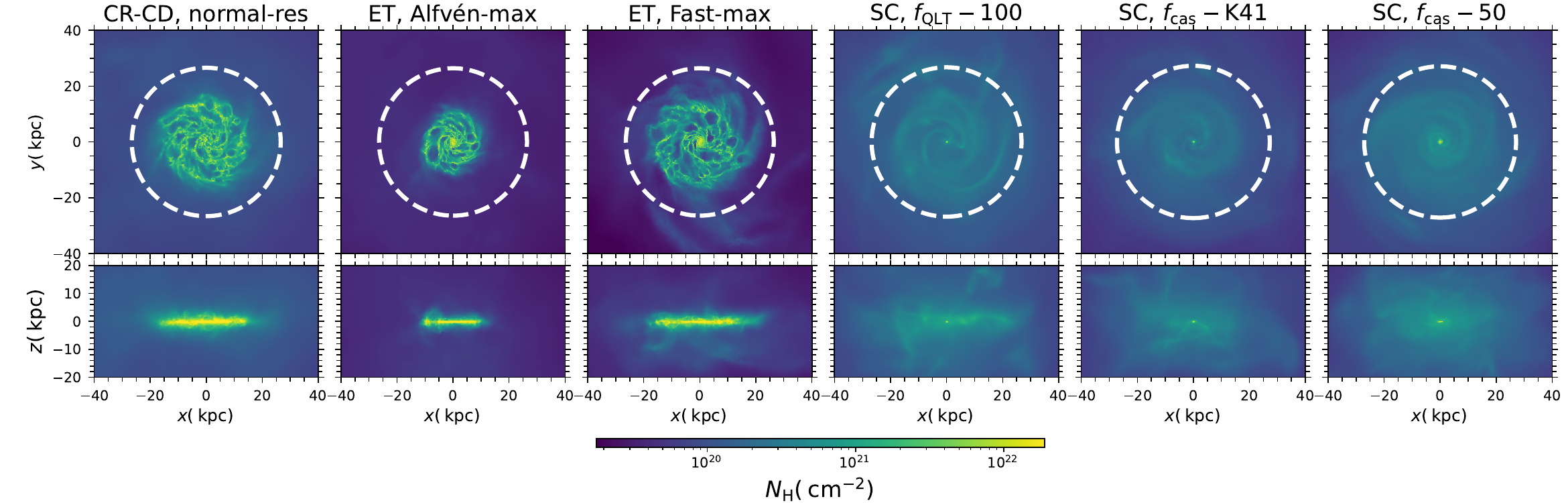}
    \caption{Similar to \fig{proj_resolution_comparison}, but for all CR runs. All maps are at $z=0$. The models (marked at the top of each set of projection maps) are listed in \tab{sims} and briefly described in \se{2.2}. All simulations are run in normal-res setup for \texttt{m12i}. Solely by eye, we can tell that the two extrinsic turbulence (ET) runs resemble the CR-CD run of the same halo much more than the self-confinement (SC) runs. This is reflected not only in the gas morphology, but also in the central stellar mass (see \tab{sims}). SC models are abnormal compared to the others, owing to the late-time SC ``blowout/runaway'' as described in \citet{ponnada2024synchrotron}.
    }
    \label{fig:proj_variant_kappa_CR}
\end{figure*}

We start the analysis with the morphological structure of the gas in each galaxy. In \figs{proj_resolution_comparison} and \figss{proj_variant_kappa_CR} we present projection maps of the hydrogen column density $N_{\rm H}$ from two directions of multiple galaxies arranged in different ways meaningful for comparisons. The projections were done with \texttt{YT} package \citep{turk2010yt}.

In \fig{proj_resolution_comparison} we compare the normal-res runs (upper half) and high-res runs (lower half) for the Hydro+, MHD+, and CR-CD runs. We see that all six runs form a disk to some extent, but the CR-CD run has the most symmetric disk with clear spiral structures. For the two non-CR (Hydro+ and MHD+) runs, increasing resolution tends to make the disk more extended with more gas accumulated around the galactic centre. Lower resolution runs also have a much more centrally concentrated stellar distribution (see \tab{sims}). At this resolution, central galaxies form too many stars by $z=0$, leaving very little gas in the CGM for late-time accretion responsible for disk formation. Our findings agree broadly with the discussion in Section 4 of \citet{hopkins2018fire}, where the resolution effects in FIRE-2 simulations were studied much more extensively. On the other hand, for the CR-CD model, the difference between the two resolutions is much less significant than those of the Hydro+ and MHD+ owing to longer-term regulation of gas accretion and star formation via non-thermal support from CRs \citep{hopkins2020but}. Furthermore, the edge-on projection maps show that the CR-CD runs have much thinner disks compared to the non-CR runs, mostly due to lower star formation rates and better alignment of the accreting material \citep{trapp2022gas, hopkins2021cosmic}. The magnetic pressure support, which is the only non-thermal support in the MHD runs, is much less important than the thermal and CR pressure supports (see details in \citet{su2018stellar,chan2022impact, ji2020properties, hopkins2020but}). 

From this simple comparison, we can see that the resolution has little effect on the basic galaxy morphology of the CR runs. However, we stress that to draw any robust conclusions on resolution effects, more detailed studies would be needed, which are beyond the scope of this work. We will briefly discuss the resolution effects again in \se{4.1.3}.

In \fig{proj_variant_kappa_CR}, we compare the gas morphology in the CR runs with different variant diffusivity models to the CR-CD run (left panel). We consider the five variant runs mentioned in \se{2.2} and \tab{sims} (the other five panels). All variant simulations are for the \texttt{m12i} galaxy, and the presented CR-CD run has normal resolution to stay consistent with other variant runs. We immediately find that while all extrinsic turbulence (ET) runs exhibit a disk structure similar to that of the CR-CD run, all self-confinement (SC) runs do not generate a well-defined disk at $z=0$ and show a very diffuse gas structure, with a very dense centre. One of the ET runs, Alfvén-max, has a smaller disk compared to the CR-CD run, but we do not find this to be a general trend, as another ET variation (Fast-max), and the same variations for a different galaxy do not show smaller disk (see \texttt{m11f} in \fig{Additional_projs}). The three SC models included here, regardless of their detailed scaling relations (see \se{2.2}), all have similar features, such as chaotic gas morphology and lack of disky structure. This is because the key of the scaling relations in the SC model is the CR scattering rate $\nu \propto e_{\rm CR}$, the CR energy density. With this relation, a higher $e_{\rm CR}$ leads to a larger scattering rate, and thus to a lower emergent effective diffusivity $\kappa$. This can lead to ``trapping'' of CRs that can locally buildup pressure and expel the ISM gas, causing a ``blowout''-like event \citep{ponnada2024synchrotron}. While such an instability is a consequence of the general feature of this CR transport method, it might not show in every galaxy. We suspect that the ``blowout/runaway'' occurs once stellar feedback-driven outflows become less efficient and after the formation of the disk, so it depends on the disk formation time. For a late-forming disk (e.g. \texttt{m11f}, see \fig{Additional_projs}), the ``blowout'' had not occurred yet. Exploring the nature of this instability in SC models and sensitivity to galaxy properties is beyond the scope of this paper and is left for future work.

In summary, the morphological structure of the disk gas, though only a very qualitative indicator of the galaxy properties, can still provide insights on the distinction between CR models \footnote{In principle, one can compare the disk cold gas mass between different runs to draw more thorough conclusions.}. In particular, the two ET models yield galaxies with well-defined disks similar to the CR-CD runs (and to the high-res Hydro+ and MHD+ runs). The ``blowout/runaway'' in the SC models is discussed further in Appendix \ref{apx:investigation}.

\subsection{Halo Gas Properties}
\label{sec:3.2}

\begin{figure*}
    \centering
    \includegraphics[trim={0.0cm 0.0cm 0.0cm 0.0cm}, clip, width =0.73 \textwidth]{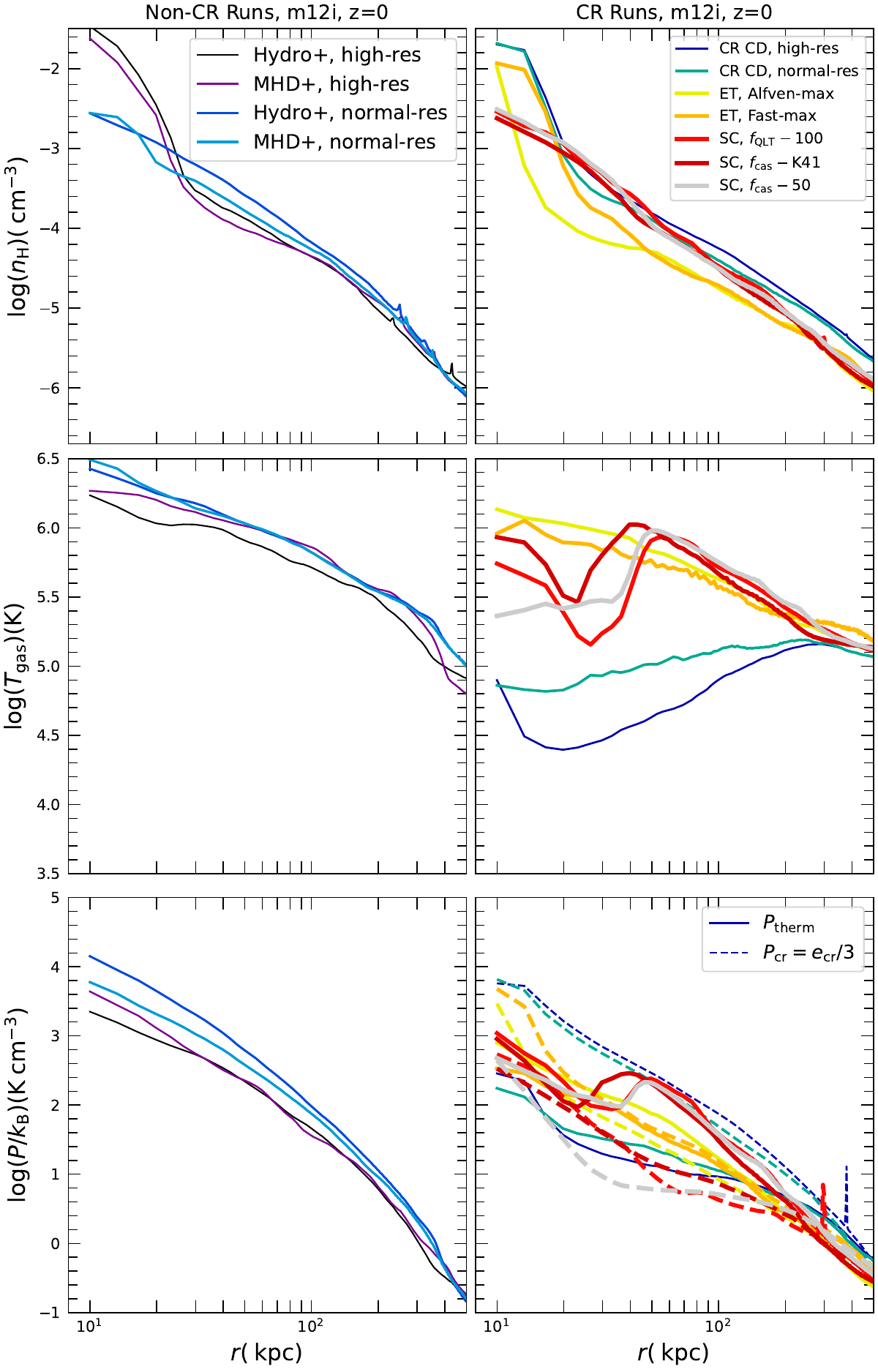}
    \caption{Radial profiles of gas number density $n_{\rm H}$ (first row), volume-weighted gas temperature $T_{\rm gas}$ (second row), and volume-weighted pressures (third row). The pressure (third) row includes both the thermal pressure $P_{\rm therm}$ (solid lines) and the CR pressure $P_{\rm cr}$ (dashed lines, for the CR runs in the right panels). We classify all of our runs into two categories as we did in \tab{sims}: the left panels are for all non-CR runs and the right panels are for all the single-bin CR runs. We label different runs are labelled with different colours and line widths to guide the eye. All runs have a target halo of \texttt{m12i}. The temperature profiles (middle row) were calculated using volume weighting, the same as was done in \citet{hopkins2021effects}, to emphasize gas cells with large volume (and corresponding high temperature). The most significant differences between different runs are seen in temperature profiles. This is consistent with what was argued in \citet{hopkins2020but} and \citet{ji2020properties} that the CRs change the temperature of the CGM gas and therefore alter its phase structure, which is critical for many of the CGM observables.}
    \label{fig:therm_prof}
\end{figure*}

\begin{figure*}
    \centering
    \includegraphics[trim={0.0cm 0.0cm 0.0cm 0.0cm}, clip, width =0.98 \textwidth]{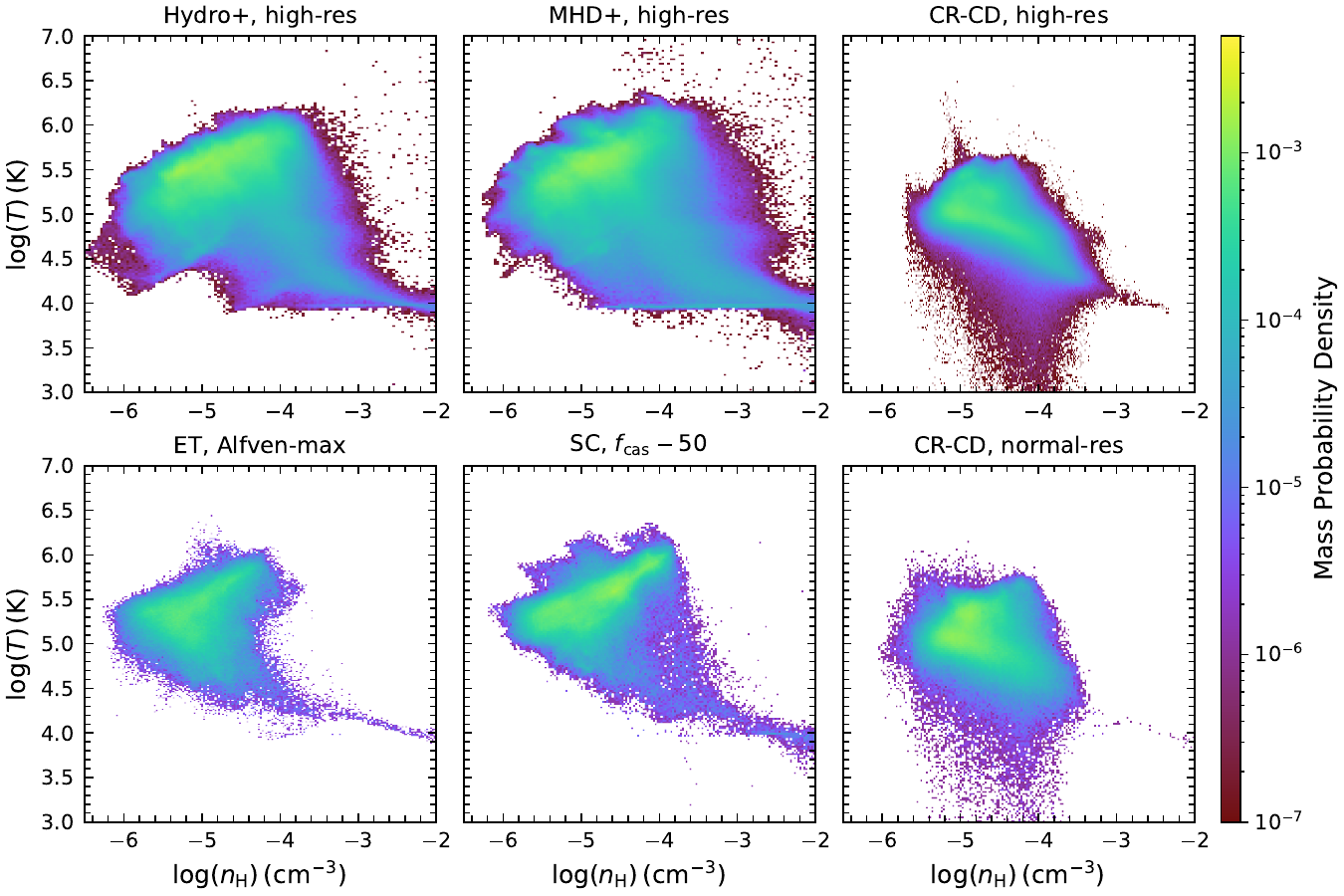}
    \caption{Mass-weighted density-temperature ($n_{\rm H}-T$) histograms of CGM gas (crudely defined as regions of $50 \kpc \leq r \leq 300 \kpc$) for six selected representative runs from our simulations (see the main text for details of selection). There is a broad resemblance among the non-CR runs and the variant diffusivity CR runs (ET and SC), while the CR-CD runs exhibit a unique phase structure. }
    \label{fig:selected_phase}
\end{figure*}


To gain more insight on how CR models affect gas properties and structures of the CGM gas, we look at the profiles of several gas properties and phase histograms of the gas surrounding our simulated galaxies. 

In \fig{therm_prof}, we plot the gas number density ($n_{\rm H}$, top row), gas temperature ($T_{\rm gas}$, middle row), and gas thermal pressure ($P_{\rm therm}$, bottom row) as functions of the three-dimensional galacto-centric radius $r$. We divide our simulations into two categories, the same as in \tab{sims}. In the bottom row, we also plot the CR pressure $P_{\rm cr}$ together with the thermal pressure in the right panel.

For the first row (gas number density), all runs appear to have a similar number density at $r\gsim 200-300 \kpc$. In the inner region $r\lsim 30 \kpc$, the difference between different runs is more visible. For the left panel, the difference between high-resolution and normal-resolution runs is the largest, as normal-resolution runs have less gas in the inner region (discussed in \tab{sims} and \se{3.1}). As for the right panel, the high-resolution and normal-resolution CR-CD runs have almost identical profiles in the inner halo, supporting our conclusion that resolution has less impact on CR runs than the non-CR runs. The two extrinsic turbulence (ET) runs also have very similar profiles compared to the CR-CD runs, but the self-confinement (SC) models deviate from other CR runs. Even at larger radii, the difference is still clear. 

The middle row (gas temperature) paints a different story. In the left panel, all the non-CR runs still behave similarly, with somewhat higher temperature in normal-res versus high-res runs. However, in the right panel, the temperature profiles diverge between different runs. The normal-res CR-CD run has significantly lower temperatures compared to the variant diffusivity runs and the non-CR runs, and the high-res CR-CD run has a temperature even lower than the normal-res run. One major reason for this is that in the non-CR runs, the primary source of pressure supporting the gas against gravity is the thermal pressure of the hot gas, but in the CR runs, the CRs are another pressure source, which supports the gas against gravity and helps reduce the requirement of the thermal pressure to support the gas, resulting in more cool gas overall. More detailed explanation can be found in \citet{ji2020properties}. However, the temperature reduction is less obvious for the ET and SC runs (see also \citet{hopkins2021effects}). The SC runs have ``peaks'' and ``kinks'' in their temperature profiles. This is expected and related to the strange morphology shown in \fig{proj_variant_kappa_CR}, and is a consequence of the ``blowouts'' mentioned previously (see Appendix \ref{apx:investigation} for more details). 

The bottom row (gas thermal pressure and CR pressure) provides a comparison between the thermal and non-thermal pressure components (mainly CR, as other non-thermal components [e.g., turbulent and magnetic] do not vary much and are typically sub-dominant; see \citet{ji2020properties} and \citet{hopkins2020but}) in the CGM. For all CR runs, the scatter between the thermal pressure profiles is not as large as in the temperature profiles because some of the temperature features are modulated by difference in density. Cosmic-ray pressure $P_{\rm cr}$, which is directly scaled as the CR energy density $e_{\rm cr}$, is quite different between different CR transport models. It is highest in the CR-CD runs and the lowest in the SC variants. Because of the non-thermal support, the corresponding temperature of the CR-dominated haloes will be lower than that of the non-CR haloes, given that they have similar total mass. This can be easily seen from the formula below for the hydrostatic equilibrium of galaxy \citep{mo2010galaxy}:

\be
    M_{\rm tot}(r)=-\frac{k_{\rm B}T(r)r}{\mu m_{\rm p}G}\left[\frac{d\ln \rho_{\rm gas}}{d\ln r}+\frac{d\ln T}{d\ln r}+\frac{P_{\rm nt}}{P_{\rm th}}\frac{d\ln P_{\rm nt}}{d\ln r}\right],
\ee

\no where $P_{\rm th}$ is the thermal pressure and $P_{\rm nt}$ is the non-thermal pressure. For CR-dominated galaxies, the extra non-thermal pressure term leads to a lower $T$ for a similar given $M_{\rm tot}$ of halo gas in approximately hydrostatic equilibrium. 

The relative contribution of non-thermal pressure is the key difference among our simulations, which in turn produce many of the trends we will see later in this paper in the CGM observables. As noted in \citet{hopkins2020but} and \citet{ji2020properties}, the dominant impact of CRs on the CGM in $L_*$ galaxies is to modify the gas temperature and overall phase structure. On the other hand, CR pressure generally tends to ``smear'' density contrasts, which is why resolution has weaker effects in CR runs. 

To further investigate how CRs can impact the gas thermal structure, in \fig{selected_phase} we show the 2D mass-weighted phase histograms for six selected runs from our simulations, including non-CR (all in high-res) runs, CR-CD (in high-res and normal-res) runs, and two chosen variant diffusivity CR runs (ET: Alfvén-Max; SC: $f_{\rm cas}-50$). The ET and SC variations exhibit little difference in their phase structures, so we only present two representative runs here. As already shown in \citet{ji2020properties}, the CR-CD run has a significantly different phase structure, with a low overall temperature and density structure compared to non-CR simulations. Due to the resolution limit, the ET and SC runs have fewer gas elements and thus less dispersion in the $n_{\rm H}-T$ phase space compared to the high-res runs, but their overall temperature is higher than the CR-CD runs, and the distribution is actually more akin to the non-CR runs. This is because the non-thermal pressure support is significantly lower in the CR-ET and CR-SC runs because of a much higher effective diffusion coefficient outside of galaxies (see \citet{hopkins2021effects}). Overall, the CR-CD run has a unique phase structure compared to all other runs. 

It is worth mentioning here that differences in CGM gas properties are \underline{not} caused by differences in star formation rates (SFRs), i.e. the recent input of CRs. In fact, most simulations with CRs have similar low redshift SFRs and SFRs in non-CR runs are also comparable. However, SFRs can be affected by CRs due to the additional non-thermal support that can modify the inflows and outflows of gas \citep{hopkins2021cosmic, trapp2022gas}. See Appendix \ref{sec:SF_histories} for a short discussion of SF histories of our simulated galaxies.

The above analysis focuses solely on the \texttt{m12i} halo/galaxy. We recognized the potential limitations from this and ran the same set of analyses for several other FIRE galaxies different from \texttt{m12i}, including CR-CD runs for \texttt{m12f}, \texttt{m12m} (both with similar halo mass to \texttt{m12i}), and \texttt{m11f}, which is an intermediate-mass dwarf. Furthermore, we also tested our results on the CR variant models of \texttt{m11f}. The temperature features of the CGM gas (one of the most important properties that controls the CGM observables) and their relative importance in runs with different CR models exhibit similar trends that we see in this section. Those are discussed in Appendix \ref{apx:addtional}.

\section{CGM Observable Constraints on CR Models}
\label{sec:Absorption} 

In this section, we attempt to constrain our CR models using observations. As many previous studies showed (e.g., \citet{ji2020properties}, \citet{hopkins2020but}), CRs are responsible for altering the phase structure of the CGM gas by changing its temperature. Hence, observables that are sensitive to CGM temperature could in principle be good reflectors of the validity of CR models. We therefore conduct comparisons for three observables, aiming to roughly represent three major gas phases: the $\HI$ absorption column for cold ($T\lsim 10^5\Kdegree$) and photoionized gas, the $\OVI$ absorption column for warm ($T\sim 2-3\times 10^5 \Kdegree$) gas, and the X-ray emission characteristics for hot ($T\gsim 5\times 10^5 \Kdegree$) gas. We present absorption column results in \se{4.1} and X-ray emission results in \se{4.2}.

\subsection{Absorption Columns}
\label{sec:4.1}

\begin{figure*}
    \centering
    \includegraphics[trim={0.0cm 0.0cm 0.0cm 0.0cm}, clip, width =0.98 \textwidth]{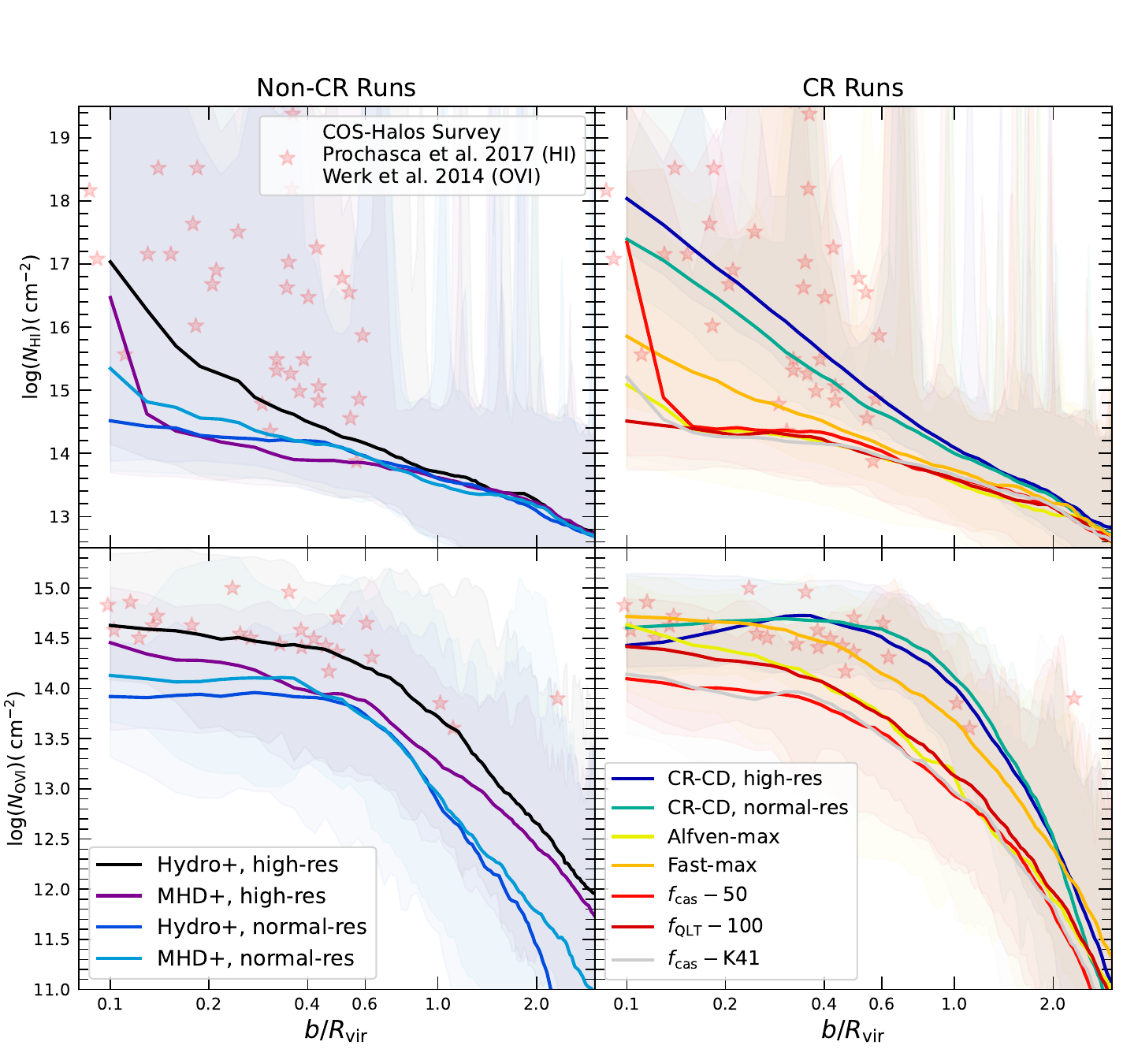}
    \caption{Column density profiles of the neutral hydrogen ($\HI$, the upper two panels) and five-times ionized oxygen ($\OVI$, the lower two panels) as functions of the impact parameter $b$ from the centre of the galaxy. We normalize $b$ with the virial radii of each run described in \tab{sims}. Similarly to \fig{therm_prof}, we put all non-CR runs on the left panels and all CR runs on the right panels. Solid lines represent the median values among all three projections (see the text for details), and the shaded regions with the same colours as the lines represent the maximum and minimum ranges. The red stars correspond to the observational data from \citet{prochaska2017cos} ($\HI$) and \citet{werk2013cos} ($\OVI$). These are all  observations from star-forming galaxies. In particular, we take data from Fig. 4 of \citet{prochaska2017cos} and $\OVI$ data for star-forming galaxies from \citet{werk2013cos}. We post-process simulations at $z\sim 0.25$ to be consistent with the observational data we are comparing, though no significant difference is seen in the results of our $z=0$ simulations (not shown in this paper).}
    \label{fig:absorption_column}
\end{figure*}

We post-process our simulation outputs and calculate the absorption column density of different ions, and then compare them to observations. We use the package \texttt{TRIDENT} \citep{hummels2017trident} to calculate the ion number densities. This requires the gas density, temperature, and metal abundances from our simulations and accounts for collisional processes and photoionization \footnote{Here, self-shielding is included, where the ionization depth equals the local Jeans length with a maximum depth of 100 pc when the self-shielding ion table is generated. }from the UV background. More details of how number densities are calculated can be found in \citet{hummels2017trident}.

Once the ion number density is calculated, we randomly choose three orthogonal directions (calling them $x$, $y$, and $z$, not necessarily the same as the axes in \figs{proj_resolution_comparison} and \figss{proj_variant_kappa_CR}) and integrate over the three directions to get the column densities. For each direction, we integrate over $\pm 4 R_{\rm vir}$ from the halo centre \footnote{However, changing this limit to any value between $4R_{\rm vir}$ to the whole simulation volume has very little effects on the final results given that those are zoom-in simulations.}. We then assign a cylindrical radial bin in each direction and take the median, maximum, and minimum values of the ion column density for each bin in all three directions. Our projection plane is a $1600\times 1600$ grid. This procedure is the same as what was done in \citet{ji2020properties}, where they compared the ion column density mainly between MHD+ and CR+ runs in Section 3.5. \footnote{We stress here that this method does not reproduce real sight lines (which in principle could be done using \texttt{TRIDENT}'s ray object), but rather, this gives more of a statistical result of the column densities. The ``true'' sight line calculation of spectral absorption features that mimics observations is computationally expensive and requires a more detailed comparison than the one in this work. However, we tested our method and confirmed that the results for the medium absorbing columns as a function of the impact parameter converge when increasing the grid numbers from our chosen resolution $1600\times 1600$. Further, compared to even very sparsely sampled ray object sight lines our projected results agree with them with $\lsim 20\%$ of discrepancy.}

When choosing the UV background models, we also follow what was done in \citet{ji2020properties}. For low ions ($\HI$, in our case), we employ the \citet{faucher2009new} (FG09 hereafter) ionization model, and for high ions ($\OVI$, in our case), we employ the \citet{haardt2012radiative} (HM12 hereafter) ionization model \footnote{Note that these are not the same as the UV background we used in the simulations.}. Details of the reasons were discussed in \citet{ji2020properties} (and a study of different UV backgrounds can be found in their Appendix B), but in short, the FG09 ionization model is more suitable for low-energy ions (e.g., $\HI$), while the FH12 ionization model is more suitable for high-energy ions (e.g., $\OVI$)\footnote{This is because the HM12 model underpredicts the inferred ionization rates of $\HI$ by a factor of $\sim 2$, whereas the FG09 model does not include both the obscured and non-obscured AGN contribution to the UV background.}.

We compare our simulation results with observations from the COS-Halos Survey \citep{werk2011cos, tumlinson2013cos}, which studies galaxies at $z \sim 0.15-0.35$. We therefore use simulation outputs at a comparable redshift, $z\sim 0.25$, and compare with column densities of $\HI$ \citet{prochaska2017cos} and $\OVI$ \citet{werk2013cos}. These are the same samples chosen by \citet{ji2020properties}. We note that while the exact values differ between our \fig{absorption_column} and \citet{ji2020properties} due to differences in projection angle and simulation output, the trends are in general agreement. While there are high-quality studies of CGM gas absorption at higher redshifts (e.g., \citet{rudie2012gaseous, liang2014mining,  steidel2018keck, chen2023cosmic}), we focus on low redshifts where CR effects are predicted to be more significant.

In \ses{4.1.1} and \ref{sec:4.1.2} we show the results for comparisons to $N_{\HI}$ and $N_{\OVI}$, respectively. In \se{4.1.3} we raise some known issues of our comparisons and discuss the resolution effects.

\subsubsection{$\HI$ Column Densities}
\label{sec:4.1.1}

In the upper panels of \fig{absorption_column} we show the column density profiles of neutral hydrogen $\HI$ as a function of the impact parameter $b$ normalized by the virial radius $R_{\rm vir}$ of each run. We compare simulation values with observational data from \citet{prochaska2017cos} in the plots. Almost all observational data points lie within the range of the maxima and minima from simulations, suggesting that simulations are consistent with observational data at first glance.

However, median values from simulations underpredict the column density of $\HI$ compared to the COS-Halos survey. This is consistent with previous work \citep{hummels2013constraints, butsky2018role} that showed that the column density of low ions in simulated CGM is significantly lower than in observations, with the problem being more severe in simulations without CRs. Note that this could also be partially caused by observational biases as HI columns in recent M31 survey are significantly lower (\texttt{AMIGA} survey, \citet{lehner2020project}). 
 
The high-res Hydro+ median line is smooth with a relatively sharp drop between the disk region ($b\lesssim 0.1 \, R_{\rm vir}$) and the CGM, and is closer to the observed columns. The high-res MHD+ run has even lower $\HI$ columns and an even sharper decline in the central region (down to $\sim 10^{14.5}\cms$, also noted in \citet{ji2020properties}). The two normal-res runs have a lower $N_{\HI}$ than observed galaxies, even around $b\sim 0.1 \, R_{\rm vir}$. This again shows that the normal-res runs have almost no gas in the CGM.

For the CR runs (upper right panel), the best match to the observation is the high-res CR-CD run. The median lines in both CR-CD runs are smooth and overlap with a set of observations, but the high-res run has higher median values of the $\HI$ columns that are in the middle of observations \footnote{\citet{butsky2023constraining} used simple model to infer lower limit of effective diffusion in the CGM and presented values higher than the adopted $\kappa$ in our CR-CD runs at very large radii. However, note that there is a large spread in inferred $\kappa_{\rm eff}^{\rm min}$ between individual absorbers, and our value is consistent with some of them, even at large radii. More importantly, while we directly compare to the $\HI$ values from simulations, \citet{butsky2023constraining} used highly uncertain (and very large) observationally inferred ionization corrections to obtain \underline{total} hydrogen column densities from the observed $\HI$ columns, making a direct comparison between these two methods rather difficult.}. This is not surprising because a larger fraction of the halo gas is in the cold/warm phase compared to non-CR runs (\citet{hopkins2020but}). The ET and SC runs underpredict the $\HI$ column density at most radii, with ``Fast-max'' version of ET showing higher median values than the rest of the variant runs.

\subsubsection{$\OVI$ Column Densities}
\label{sec:4.1.2}
In the lower panels of \fig{absorption_column} we show the column density profiles of the five-times ionized oxygen ($\OVI$). Although the nature and origin of $\OVI$ in the CGM is still unclear, there are several hypotheses including collisional ionization of virial-temperature gas \citep{oppenheimer2016bimodality}, photoionization due to flickering AGN \citep{oppenheimer2018flickering}, and turbulent mixing layers \citep{fielding2020multiphase}.
Previous work \citep[e.g.][]{ji2020properties} also showed that the inclusion of CRs significantly changes the temperature structure, which can affect both the collisional ionization and photoionization processes that are responsible for the production of $\OVI$. Therefore, the absorption column of $\OVI$ could be an important factor to use when constraining the CR models.

The values of $N_{\OVI}$ for each run roughly follow the same order as for $N_{\HI}$. For the non-CR runs (lower left panel in \fig{absorption_column}), only the high-resolution Hydro+ run has $N_{\OVI}$ close to the observations, which are higher than the values in other non-CR runs. There are several reasons for this trend. On the one hand, the temperature of the halo gas is the lowest amongst non-CR runs in the high-resolution Hydro+ simulation (see \fig{therm_prof}), while the abundance of oxygen in its halo is slightly higher in the normal-resolution run (not shown). This indicates that temperature plays a slightly more important role here. On the other hand, it is also possible that high-resolution simulations better resolve smaller clumps of gas and their interaction layers with the diffuse gas (more detailed discussion in \se{4.1.3}). Since CRs lower the temperature of the CGM gas, CR runs have slightly higher $N_{\OVI}$ median values, as expected. This is because the temperature at which $\OVI$ is the most prevalent ion of oxygen is $ \sim 2-3\times 10^5\Kdegree$ for collisional ionization, and the non-thermal pressure support from CRs lowers the halo gas temperature to make it closer to this range. Both high-resolution and normal-resolution CR-CD runs, as well as the ``Fast-max" ET run, are in good agreement with observations. However, all other variant runs have $N_{\OVI}$ well below the observed values in the CGM, but still slightly higher than most non-CR runs.

\subsubsection{Known Issues and Resolution Effects}
\label{sec:4.1.3}

Our simulation comparison has a few caveats, but we believe that the analysis holds despite these minor shortcomings. We discuss these caveats here in detail, but we will also recapitulate in \se{5.1}.

One of the potential issues is the broad range of halo masses in the COS-Halos survey compared to a single simulated halo (with relatively high mass). At $M_{\rm vir}\sim 10^{12} M_{\odot}$ the halo gas in our simulated haloes has just transitioned to a virialized hot halo and its inner CGM also virializes \citep{kerevs2005galaxies, dekel2006galaxy, kerevs2009galaxies, stern2019cooling, stern2021virialization}, with details also dependent on feedback from cosmic rays \citep {ji2021virial}. Because of this, at a slightly higher redshift and a slightly lower halo mass, the CGM is expected to contain a higher fraction of cold gas, increasing the column density of HI \citep{kakoly2025turbulence}. At the same time, at similar halo masses, the properties of galactic winds that drive strong outflows are expected to change significantly \citep{muratov2015gusty, yu2021bursty}. One therefore needs a more complete sample of simulated halo masses and redshifts to compare more carefully to the observed samples. Nevertheless, it is still meaningful to see the differences between simulated galaxies. Overall, there is a broad trend that CR runs (especially high-resolution CR-CD) better match the observed range of CGM absorbers.

Another important factor that can significantly change the simulated CGM is the resolution. In the past decade or so, there has been a significant tension between simulations and observations of the low-$z$ CGM, because simulations underpredicted various ions observed by COS-Halos and other surveys. Recent computational efforts have focused on utilizing finer numerical resolution throughout the simulated CGM in order to relieve this tension \citep{ 2016MNRAS.461L..32F, hummels2019impact, 2019MNRAS.482L..85V, peeples2019figuring, ramesh2024zooming, hummels2024cloudflex}. These studies have demonstrated that finer numerical resolution can increase the proportion of cool small-scale gas clouds, thus bringing simulations more in line with observations of $\HI$ and other low ions. Such a trend is seen in our simulations on the left panels of \fig{absorption_column}, but in our case better resolved clouds are only a \textit{secondary} effect, as most of the differences in the cold halo content with resolution come from the different efficiency of feedback and galaxy formation that dramatically change the remaining gas fraction of the CGM. Another potential effect from increased resolution is that turbulent mixing layers of small clouds that form at high resolution could also boost the availability of a moderate temperature $\OVI$-bearing gas
and boost the OVI absorption (e.g., \citet{gronke2023cooling}), but there is no concrete evidence supporting this effect in results presented in this paper as such mixing layers are not resolved in our simulations, even in the high-resolution runs. We leave a detailed study of the cloud structure of the non-CR runs to future work.

Incidentally, our high-res Hydro+ run of \texttt{m12i} exhibits very non-uniform distribution and a relatively strong time variation of {\OVI} in the CGM (not shown), partially causing the medium values of $N_{\rm OVI}$ to be significantly higher than the MHD version of the same halo. However, we ran the same pipeline to the high-res Hydro+ runs of two additional MW-mass FIRE galaxies (\texttt{m12f} and \texttt{m12m}), and their $N_{\rm OVI}$ profiles are more akin to the MHD runs' (not shown). We therefore assert that the \texttt{m12i}'s high-res Hydro+ run is anomalous compared to our other MW-mass haloes.

Despite these potential effects, it is notable that numerical resolution does not appear to have a major impact on the CR runs. There is a mild boost in the $\HI$ content in the CGM between the normal-res and high-res CR-CD runs, and there is virtually no difference in the $\OVI$ content of these haloes. This further suggests that the physical mechanisms responsible for the production of cool ($\HI$) and warm ($\OVI$) gas can differ between the CR and non-CR simulations. Previously, \citet{mccourt2018characteristic} suggested that simulations would require spatial resolution thresholds $< 1 \pc$ or explicit subgrid models \citep{butsky2024galactic, hummels2024cloudflex} for accurate treatment of thermally unstable gas in the CGM. However, with CRs, the extra non-thermal pressure modifies the formation of small clouds \citep{2020ApJ...903...77B}, which suggests that the CGM in current simulations might not need such stringent resolution requirements when CRs are included. 

\subsection{X-ray Emission}
\label{sec:4.2}

Massive galaxies (much more massive than the Milky Way) are expected to be surrounded by hot virialized gas ($T\gg 5\times 10^5 \Kdegree$) that emits thermal free-free and metal-line cooling radiation at X-ray frequencies \citep{white1978core,snowden1998progress,kerevs2005galaxies, henley2013xmm}. This hot gas is observed routinely in galaxy clusters, but at lower masses (e.g., Milky Way), the CGM of individual galaxies remains undetected -- as expected given their much lower virial temperatures -- although there have been some detections of emission in the inner CGM, close to galaxies \citep{li2013chandra} or from larger radii in stacks of large galaxy samples \citep{zhang2024hot}. On the other hand, future mission concepts have sought to use X-ray emission in MW-mass systems as a direct constraint on strong winds and feedback (which could generate X-ray emission in such low-mass, lower-temperature halos through strong shocks; see \citet{2016MNRAS.463.4533V, 2022arXiv221109827K}). CRs could therefore strongly modify the X-ray emission properties through a variety of processes: CR-driven winds generating shocks, CR pressure leading to modified CGM temperatures and densities (as discussed above), or -- as pointed out in \citet{hopkins2025cosmicraysmasqueradinghot} -- CRs directly contributing to X-ray emission through non-thermal processes like inverse Compton scattering (ICS) of background radiation.

In this section, we estimate the emitted X-ray luminosity $L_{\rm X}$ from our simulated galaxies and show how $L_{\rm X}$ depends on stellar mass $M_{*}$ and the star formation rate (SFR). We then briefly discuss the observation samples we selected in \se{4.2.1} and show our comparison results in \se{4.2.2}. 

We calculate the X-ray thermal cooling luminosity for each gas particle by interpolating an X-ray emission table computed from the Astrophysical Plasma Emission Code (APEC, v3.0.9; \citet{smith2001collisional}; \citet{foster2012updated}), assuming the gas is optically thin and in collisional equilibrium. This procedure accounts for X-ray from both thermal free-free (bremsstrahlung) and metal line emission. For non-thermal emission, we follow \citet{hopkins2025cosmicraysmasqueradinghot} by assuming -- motivated by direct observations of CR spectra in the Solar system -- that most of the CR lepton energy must escape the galaxy (since the energy is primarily in $\sim\,$GeV CR electrons with lifetimes $\sim 10^{9}\,$yr and residence/escape times in the Galaxy of $\sim 10^{7}\,$yr; e.g. \citealt{dimauro:2023.cr.diff.constraints.updated.galprop.very.similar.our.models.but.lots.of.interp.re.selfconfinement.that.doesnt.mathematically.work}), which lose their energy in the CGM primarily to IC scattering of CMB photons to $\sim 1$\,keV energies. Per \citet{hopkins2025cosmicraysmasqueradinghot}, assuming a local ISM-like CR proton and electron spectrum, and integrating over the total lifetime of the CRs, this produces a soft X-ray IC luminosity $L_{\rm X,\,IC} \sim 0.02\,\dot{E}_{\rm cr}$ where $\dot{E}_{\rm cr}$ is the total (galaxy-integrated) CR injection rate into the CGM. Here $\dot{E}_{\rm cr} \approx 10^{50}\,{\rm erg}\,\dot{N}_{\rm SN}$ in terms of the SNe rate $\dot{N}_{\rm SN}$\footnote{Note that this simply comes from the assumption that $10\%$ of SN explosion energy becomes CRs, and a single SN releases $10^{51} {\rm \, erg}$ energy, consistent with the assumption in the simulation.}, which we take from the same rates used in code (averaged over the last $\sim 1\,$Gyr, the CR lifetime). This method practically captures the upper bound of  the true ICS emission. However, because we do not evolve the full spectra of CRs (see \se{2.2}), we simply do not have enough information to truly model the ICS. Nevertheless, the true values should not be significantly affected by the choice of the loss factor or the inclusion of full physics, as discussed in \citet{hopkins2025cosmicraysmasqueradinghot}. 

At $z=0$, local increase of star formation in FIRE galaxies does not result in large-scale global outflow from galaxies, but instead forms hot superbubbles, especially at high resolution (not shown here but see \citet{chan2022impact}). Those superbubbles reside near the disk and can significantly affect the total X-ray emission from galaxies. If the hot gas dominates the CGM, contribution from superbubbles is typically small in MW mass galaxies. However, in the CR-CD runs, halo gas far from galaxies does not produce much thermal X-ray emission, so these intermittent superbubbles could render strong time variability to our results. To illustrate such time variability of the thermal X-ray emission, we show the range of X-ray emission from 5 snapshots for the CR-CD simulations, presented by the ``error bars'' in the figures. The included snapshot range corresponds to $\sim 200 \Myr$ from $z=0$, which is much longer than the lifetimes of individual superbubbles.

\subsubsection{Different Samples and Selection}
\label{sec:4.2.1}

We compare our simulations with three sets of observational data: \texttt{ROSAT} data from \citet{anderson2015unifying}, \texttt{eROSITA} data from \citet{zhang2024hot}, and \texttt{Chandra} data from \citet{li2013chandra}. The first two focus on {\it stacked} data from surveys of over 250,000 galaxies, enabling detection of diffuse soft X-ray emission from the CGM out to $\gtrsim 100\,$kpc from galaxies. In contrast, \citet{li2013chandra} focuses on a few individually detected bright nearby X-ray sources (e.g., M82) and the emission from the very near CGM ($\lesssim 10\,$kpc) outside the disk. Various spatial cuts were applied in all the works to exclude inner regions of their galaxies \footnote{Some of them are the same comparison targets as in \citet{chan2022impact} (Fig. 12). However, in that work, the authors applied a rather crude spatial cut for all simulated galaxies by only including gas cells within $10\kpc$ from the centres of the galaxies.}. In this work, we follow what was done in these three observational papers individually to better mimic their real spatial cuts. In Appendix \ref{apx:addtional} we include additional simulations and then compare them with the same data sets.

\begin{figure*}
    \centering
    \includegraphics[trim={0.0cm 0.0cm 0.0cm 0.0cm}, clip, width =0.98 \textwidth]{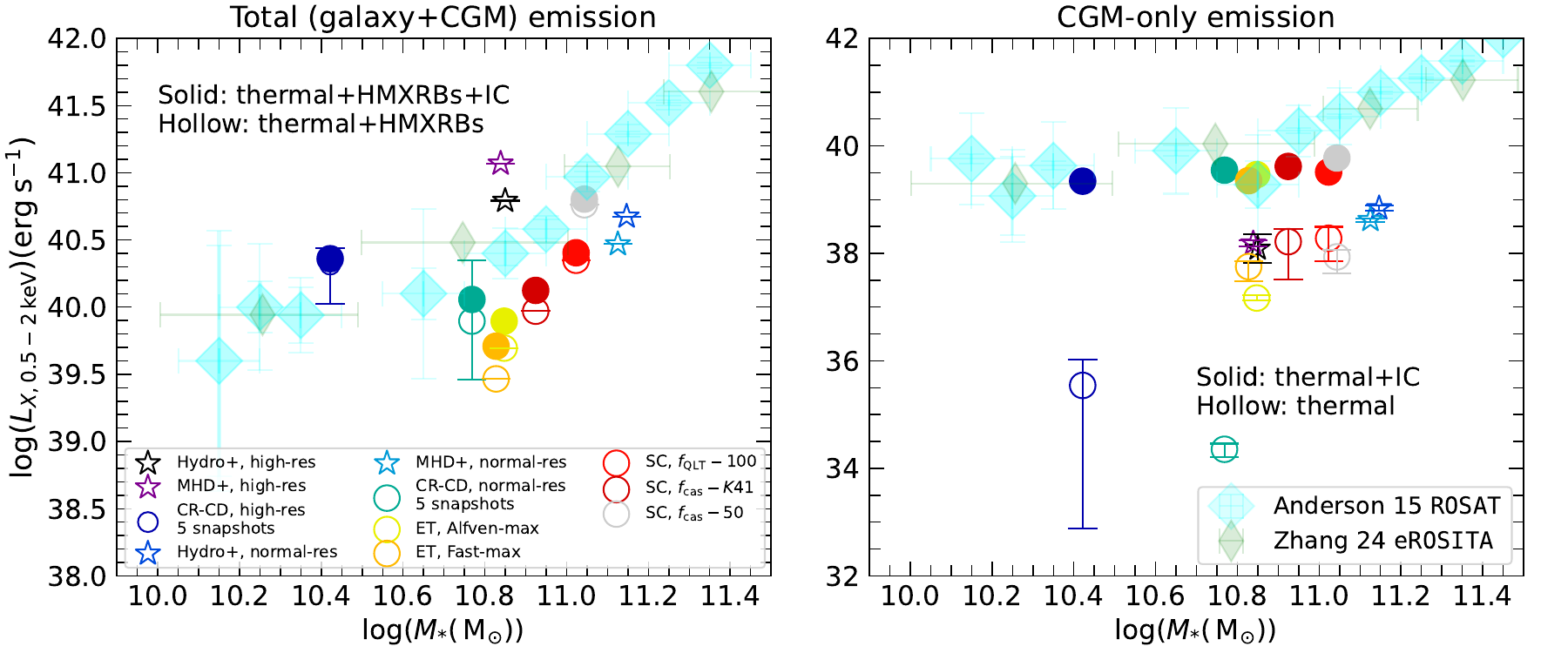}
    \caption{Soft X-ray luminosity $L_{\rm X}$ versus stellar mass $M_*$. We compare stacked \texttt{ROSAT} (\citet{anderson2015unifying}, cyan diamonds with error bars) and \texttt{eROSITA} (\citet{zhang2024hot2}, green thin diamonds with error bars) data to simulations with similar cuts mimicking observations (see \se{4.2.1}). Coloured shapes (both hollow and solid) show simulations as labelled. For each simulated galaxy, the ``error bars'' here show range of X-ray luminosity for 3 different projections; for CR-CD runs the luminosity range also includes variations over the past $200 \Myr$ (see the main text). Note that two panels have different ranges in the $y$-axes. \textit{Left:} Total emission including central galaxy and CGM. The hollow shapes show the thermal emission and HMXRBs (these contribute comparably to hot ISM gas, see the main text). For the solid shapes we add ICS emission (only for the CR runs) from CRs in the CGM scattering CMB photons to $\sim 1\,$keV energies, but it is small compared to total thermal ISM+HMXRBs. \textit{Right:} CGM emission subtracting central galaxies and point sources. Similar to the left panel, we show results including the contribution from ICS (solid shapes) separately from the thermal-only results (hollow shapes). For simplicity we assume that all of $L_{\rm X,\,ICS}$ is radiated in the CGM. $L_{\rm X,\,ICS}$ is predicted to dominate the soft X-ray CGM emission in the simulations with CRs, compared to the thermal emission. }
    \label{fig:X_ray_Anderson}
\end{figure*}

\begin{figure}
    \centering
    \includegraphics[trim={0.0cm 0.0cm 0.0cm 0.0cm}, clip, width =0.48 \textwidth]{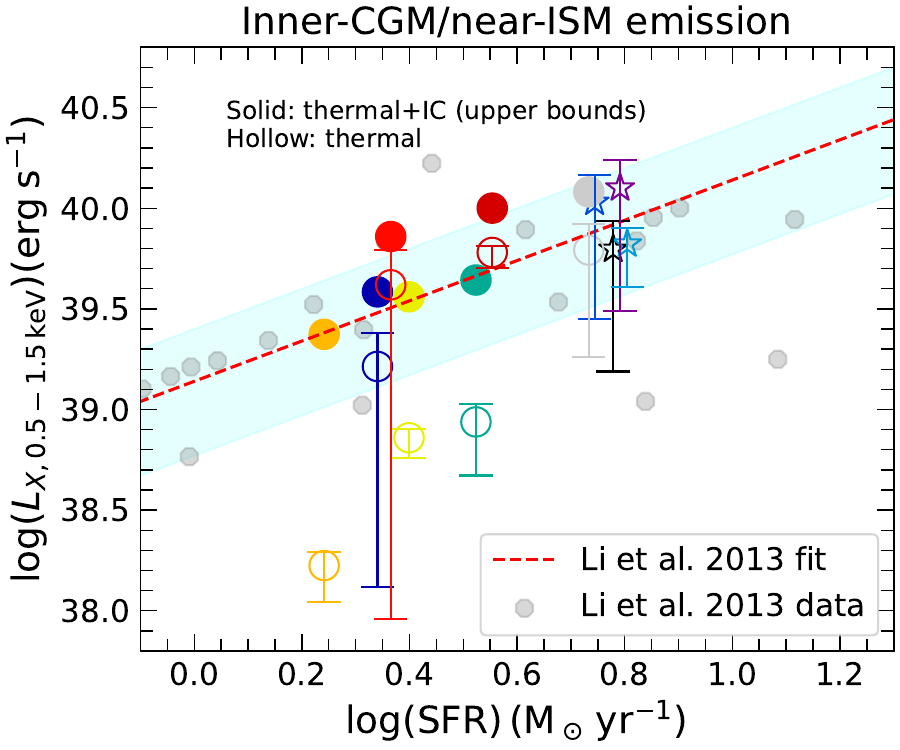}
    \caption{Soft X-ray luminosity in the ``inner-CGM/near-ISM'' region ($\sim 1-10 \kpc$ above/below the disk) of individual X-ray bright nearby star-forming galaxies from \citet{li2013chandra}, as a function of SFR. Coloured shapes show thermal-only (hollow) and thermal+ICS (solid) $L_{\rm X}$ from these regions with spatial cuts described in \se{4.2.1} and are labelled the same way as \fig{X_ray_Anderson}. ``Error bars'' here again represent the ranges from different projection directions. The median relation is consistent with $\sim 20-50\%$ of the ISM+XRBs emission in \fig{X_ray_Anderson}, which is expected given that this is defined as the X-ray luminosity just outside one fitted X-ray scale-height of the disk. }
    \label{fig:X_ray_Li}
\end{figure}

\citet{anderson2015unifying} stacked \texttt{ROSAT} data around galaxies of different stellar masses, and presented both the ``total'' ({\it including} the central galaxy and all point sources like AGN and resolved X-ray binaries [XRBs]), and their estimate of ``CGM-only'' X-ray emission (subtracting those sources, as the extended emission around lower-mass systems is not resolved in \texttt{ROSAT}). For comparison with the total emission, we sum the thermal emission from all gas cells within a projected impact parameter $r<2R_{\rm 500}$ along three random directions (here, $R_{500}$ is defined relative to critical) and take the averages,\footnote{Since most of the emission in this sample comes, as observed, from a combination of ISM+XRBs, there is extremely small impacts on the final results if we change this limit to anything between $R_{\rm vir}$ (or $R_{\rm 500}$) and $\infty$.} plus the total $L_{\rm X,\,IC}$ predicted, plus XRBs. XRBs are not evolved explicitly in-code, so we add LMXB and HMXB emission using the fitting functions as a function of stellar mass and star formation rate from \citet{anderson2015unifying}, which here are completely dominated by HMXBs (their Eq.~E2) with $L_{\rm XRB} \approx 1.4 \times 10^{39}\,{\rm erg\,s^{-1}}\,\dot{M}/({\rm M_{\odot}\,yr^{-1}})$. For comparison with the ``CGM-only'' emission, we take the average thermal emission in a projected impact parameter $0.15 R_{500}<r<R_{500}$ (the region \citet{anderson2015unifying} heuristically assigns to this emission, based on more massive resolved halos). For both total and CGM-only emission, we add total IC luminosity on top of the thermal emission using the method described at the beginning of \se{4.2}. This set of observational data and our calculated $L_{\rm X}$ are shown in \fig{X_ray_Anderson}.

More recently, \citet{zhang2024hot,zhang2024hot2} stacked soft X-ray ($\sim$\,keV) emission from \texttt{eROSITA} and were able to detect the extended halos around MW-mass systems out to $\gtrsim 100\,$kpc (showing that most of the diffuse CGM emission comes from these larger radii). We show their data for both the total and the CGM-only emission (specifically their point-source, background, central-and-satellite galaxy subtracted luminosities) alongside that from \texttt{ROSAT} in \fig{X_ray_Anderson}.

Finally, we compare in \fig{X_ray_Li} with harder \texttt{Chandra} X-ray emission in the central $\sim 1-10\,$kpc -- i.e.\ the ``inner-CGM/near-ISM'' region -- around the few X-ray brightest low-mass star-forming galaxies, from \citet{li2013chandra}. The authors excluded the central galaxy either by excluding a circular mask with $R=R_{25}$ (the radius where the optical surface brightness falls below $25\,{\rm mag \, arcsec^{-2}}$, for face-on systems, which are the minority of the sample) or a rectangular mask with length $\pm R_{25}$ along the disk and height $-h_{-}$ to $+h_{+}$ (where $h_{\pm}$ are the scale heights given by fitting the mean X-ray emissivity profiles above/below the disk to exponential functions). We mimic this by projecting our galaxy along three perpendicular directions, calculating vertical $L_{X}$ profiles from the cooling luminosity and fitting to $h_{\pm}$, and taking $R_{25} \sim 0.05\,R_{\rm vir}$ to approximate the optical radius (as found in observations; see  \citet{kravtsov:2013.size.vs.rvir.relation}).\footnote{Choosing $R_{25} \sim 0.015,\,0.025,\,0.05$ give similar results here.} Since in \citet{li2013chandra} all X-ray point sources were resolved and removed, we did not include any XRBs here. For $M_*$ shown in \fig{X_ray_Anderson} we use the values from \tab{sims} and for the SFR shown in \fig{X_ray_Li} we account for the stars formed within $100 \Myr$. \footnote{Note that here stellar mass loss is not accounted, but it should be negligible for our analysis.}

Note that the emission \citet{li2013chandra} detected is observed to come from just a couple of kpcs above/below the disk, while most of the IC comes from much larger radii $\sim 100\,$kpc (\citet{hopkins2025cosmicraysmasqueradinghot}), like the extended soft X-ray halos \citep{zhang2024hot}. We therefore stress that the calculated $L_{\rm X}$ from simulations with the contributions from ICS included serves as an \textit{upper bound} to the resulting emission.

\subsubsection{Results}
\label{sec:4.2.2}

In \fig{X_ray_Anderson}, we see (consistent with \citealt{anderson2015unifying}) that most of the ``total'' X-ray emission comes from the central galaxy ISM+XRBs. Note that \citet{anderson2015unifying} argued that it could all come from resolved XRBs, while some other works (e.g., \citet{Mineo2012,Mineo2014}) argued that $\sim 2/3$ of the central $L_{X}$ comes from resolved XRBs in the nearest star-forming galaxies. Here we find that the contribution of hot ISM gas is comparable to that of XRBs in a broad sense but often larger (not presented in the figure). This is expected: for star-forming galaxies the XRB luminosity is dominated by short-lived HMXBs with $L_{\rm XRB} \sim 1.4 \times 10^{39}\,{\rm erg\,s^{-1}}\,\dot{M}_{\ast}/({\rm M_{\odot}\,yr^{-1}})$, while the SNe rate is dominated by core-collapse SNe with 1\,SNe per $\sim 100\,M_{\odot}$ formed. If we assume a fraction $\epsilon_{X,\,\rm SNe}\sim \epsilon_{X,\,\rm SNe}^{0.01} 0.01$ (motivated by detailed blastwave calculations; \citealt{chevalier:1974.sne.breakout.conditions}) of SNe energy is thermalized and radiates in soft X-rays before it adiabatically cools, then the ISM hot gas luminosity from SNe shocks should be $L_{\rm X,\,ISM} \sim 3 \times 10^{39}\,{\rm erg\,s^{-1}}\,\epsilon_{X,\,\rm SNe}^{0.01}\,\dot{M}_{\ast}/({\rm M_{\odot}\,yr^{-1}})$.

Comparing \fig{X_ray_Li} with both panels of \fig{X_ray_Anderson}, we see that the ``inner-CGM/near-ISM'' extended emission as defined by \citet{li2013chandra} tends to be an $\mathcal{O}(1)$ fraction of the ``total'' (galaxy ISM+XRB, left panel of \fig{X_ray_Anderson}) luminosity, typically $\sim20-50\%$, and comparable to the ``CGM-only'' emission (right panel of \fig{X_ray_Anderson}) in both observations and simulations. This follows immediately from the definitions in \citet{li2013chandra}: for most of the galaxies in the sample (which are edge-on) the X-ray emission from the galaxy is fitted to a vertical exponential profile, and emission at $< 1$ scale height $h$ is attributed to the galaxy while that at $>1\,h$ is considered ``extended''. For a vertical exponential, this gives a fraction $=0.36$ of the total emission in the ``extended'' component. The only simulations which appear to fall low on this relation in their thermal emission are the same which fall low in total thermal-only emission in \fig{X_ray_Anderson} (e.g. the ET runs), which then are boosted by the inclusion of ICS. Because ICS dominates mainly in the extended CGM that is detectable only in stacked observations, X-ray emission from simulations must include ICS to match observed extended CGM emission; by contrast, thermal X-ray emission is in agreement with individual inner-CGM observations with or without addition of ICS. However, an important caveat here is that the \citet{li2013chandra} sample specifically selected X-ray bright nearby star-forming galaxies: it is {\it not} a SFR or mass or volume-limited sample (unlike the \citealt{anderson2015unifying} and \citealt{zhang2024hot2} samples).

In diffuse and extended CGM emission (right panel of \fig{X_ray_Anderson}), we see that when accounting for the expected ICS emission, the CR runs appear broadly consistent with the observations, as argued from more general empirical grounds in \citet{hopkins2025cosmicraysmasqueradinghot}. Note that we include the expected CR-ICS \textit{only} for the CR runs. Some of the most-massive non-CR or SC runs ($M_{\ast} > 10^{11}\,M_{\ast}$) remain slightly low. This may be because we neglect AGN feedback (which could boost the thermal and CR-ICS emission in more massive galaxies). More strikingly, the predicted ICS emission dominates over the thermal emission for these low-mass halos (as predicted in \citet{hopkins2025cosmicraysmasqueradinghot}), with the thermal/cooling diffuse CGM emission generally a factor of $\sim 30-100$ below the observations. This is consistent with the behaviour of other large-volume cosmological simulations (e.g. TNG, EAGLE, FLAMINGOs, BAHAMAs, MAGNETICUM, as shown in \citealt{truong:2023.cosmo.sims.sb.predictions.verylow.larger.vs.erosita.data,zuhone2024propertieslineofsightvelocityfield, silich2025x, grayson2025hot, shreeram2025retrieving,lau2025x}), and constraints from X-ray absorption studies (which give upper limits to the mass of hot, metal-rich gas at $R\gtrsim 100\,$kpc incompatible with the observed soft X-ray emission if it were all of thermal origin; see \citealt{yao:2010.chandra.upper.limits.warm.hot.cgm.gas.in.mw.mass.halos,ponti:2023.erosita.supervirial.gas.close.to.galaxy.cgm.low.metal.and.low.density}), as well as cosmological parameter and total galactic metal-budget constraints (see references in \citet{hopkins2025cosmicraysmasqueradinghot}). 

This supports the \citet{hopkins2025cosmicraysmasqueradinghot} argument that CR-ICS cannot be neglected in soft X-ray diffuse extended emission around MW/M31-mass galaxies. If so, the {\it spatially-resolved} properties of this emission can provide direct constraints on CR propagation, motivating more detailed comparisons in future work.

Note that in both total and CGM emission, our high-resolution CR-CD simulation produces larger $L_{\rm X}$ than the normal-resolution CR-CD run, despite it having a lower stellar mass, but also shows more variability in CGM cooling emission, owing to the lower temperatures where emission is more sensitive to clumping and shocks (\se{3.2}). Thus, resolution effects and time variability could still be significant here. 

\section{Conclusions and Outlooks}
\label{sec:Conclusions}

\subsection{Conclusions}
\label{sec:5.1}

We study the CGM properties and observables of FIRE-2 simulations of a Milky Way (MW)-mass galaxy (\texttt{m12i}) evolved with different CR transport physics. 
Our main goal is to constrain CR transport models and study CR influence on galaxy evolution. Our main conclusions are as follows. 

\begin{enumerate}
    \item \textbf{CGM morphology and resolution effects: }Most \texttt{m12i} runs have a clear disk, as shown in multiple previous FIRE papers. However, for the non-CR runs, the mass resolution affects the morphology more significantly than the CR runs, with the normal-resolution runs running out of gas in the galaxy and the CGM. Excessive star formation at earlier times and less efficient stellar feedback are the main cause of this \citep{hopkins2018model}. However, in the CR-CD runs, non-thermal support sustains gas inside the galaxy and in the CGM even at normal resolution, resulting in much less difference between the high- and normal-resolution runs. As for our variant CR runs (all in normal-resolution), the ET models exhibit a clear disky structure similar to the usual \texttt{m12i} runs, but the SC models show almost no coherent disk at $z=0$. This is due to an early ``runaway/blowout'' event in the galaxy. This blowout occurs at slightly different epochs around $z\sim 0.2$ for \texttt{m12i} simulations. 
    In general, the self-confinement (SC) models are prone to strong instability in disk galaxies because of the scaling of CR diffusion in SC runs. The exact timing could be sensitive to various factors, including halo mass, mass resolution, model parameters, and star formation/halo evolution histories.

    \item \textbf{Thermal properties: }All \texttt{m12i} runs exhibit similar density profiles due to the same host halo. The difference between temperature profiles can be significant, which creates a critical difference in the phase structure of the CGM, resulting in distinctive CGM properties. The CR pressures are higher than the thermal pressures for both the high-res and normal-res CR-CD runs, but the two pressure components are more comparable for the ET runs, in agreement with the equipartition theorem. Due to the ``blowout'' events, the SC runs have additional features in both temperature and pressure profiles, making them difficult to interpret.

    \item \textbf{Comparisons to observations: }We post-process our simulation data to obtain the absorption column densities of $\HI$ and $\OVI$ and X-ray emission luminosities and compare them to observations. We present our conclusions from different comparisons below.
    \begin{itemize}
        \item \textit{Ion absorption columns:} By comparing the low-$z$ star-forming galaxies from the COS-Halos survey to our single simulated galaxy \texttt{m12i}, we find that the non-CR runs (Hydro+ and MHD+) tend to underpredict both median $N_{\HI}$ and $N_{\OVI}$ compared to observations. Increasing resolutions helps boost the ion content, helping high-resolution Hydro+ run to be a good match to the observed $N_{\OVI}$, but it still underpredicts the $N_{\HI}$. The CR-CD runs predict both $N_{\HI}$ and $N_{\OVI}$ more consistent with observations, due to their reduced CGM gas temperature that boosts the $N_{\HI}$, and, to a smaller effect, the $N_{\OVI}$. The variant diffusivity CR runs (including both ET and SC models) in general underpredict both $N_{\HI}$ and $N_{\OVI}$, with potential exception of Fast-Max ET run that is in reasonable agreement with observed $N_{\OVI}$. Of all runs we tested, the CR-CD runs are still in the best agreement with observations. Overall, the resolution change has stronger impact on the non-CR runs than the CR runs. Other limitations might affect our conclusions, which we discuss further below in \textit{Caveats and limitations of our comparisons}.
        
        \item \textit{X-ray emission:} We applied different cuts to our simulations to mimic different observational samples. In a comparison to stacked emission from large galaxy surveys in \texttt{ROSAT} and \texttt{eROSITA} (\citet{anderson2015unifying} and \citet{zhang2024hot, zhang2024hot2}) {\it including} the central galaxy and point sources (XRBs), the total emission from our simulated galaxies is in reasonable agreement with the observations (with comparable thermal SNe-heated ISM, and XRB emission), with and without including ICS emission. On the other hand, the extended ($\sim 100\,$kpc) CGM emission appears to agree well with the predicted soft X-ray emission dominated by the IC scattering of CMB photons by low-energy ($\sim 0.1-1\,$GeV) CR electrons in our CR simulations, as predicted in \citet{hopkins2025cosmicraysmasqueradinghot}, while the thermal (free-free+metal-line) cooling emission from these large radii is much smaller than observations in all of our simulations (consistent with most other MW-mass galaxy simulations). The ``inner-CGM/near-ISM'' X-ray emission detected at $\lesssim 10\,$kpc from nearby resolved star-forming galaxies (e.g., M82) in \citet{li2013chandra} appears consistent with the simulated thermal emission of most of our simulations, with the exception of ET variants and normal-res CR-CD simulations, but the inclusion of ICS boosts up the values and makes all simulations consistent with the range of observations (this sets the upper bound for the X-ray emission as it assumes total ICS emission originates in the selected spatial region).
        
        \item \textit{Caveats and limitations of our comparisons:} We acknowledge that our comparisons have several drawbacks: \textit{(1). Incompatibility between our simulated galaxies and observed samples.} This applies to both our ion absorption and X-ray comparisons. In regard to the ion absorption comparison, the COS-Halos survey covers a wider range of halo masses with focus on galaxies with smaller masses ($M_{\rm halo}\lsim 10^{12}\Msun$), which is a key difference with our simulations of a single halo at $M_{\rm halo}\sim 10^{12}\Msun$ (see \se{4.1.3}). For these reasons, our comparisons are not fully ``apples-to-apples'', so we focus mostly on the run-to-run variations. In regard to the X-ray comparison, as mentioned in \se{4.2}, galaxies from large surveys like \texttt{ROSAT} and \texttt{eROSITA} are stacks and are not classified by types, and thus may not necessarily be a good match to our individual galaxies. However, other large-volume simulations (e.g., TNG, EAGLE) give results broadly similar to ours in a sense that their thermal X-ray emission is lower than the total observed X-ray emission in the MW-mass regime \citep{zhang2024hot2}. When comparing to individual observed galaxies \citep{li2013chandra}, the scatter of the data is not small enough to distinguish between different CR models, especially given the complicated selection effects. \textit{(2). Resolution effects.} The variant diffusivity CR runs have limited resolution, and some of them have already been improved in FIRE-3 (e.g., \citet{ponnada2025hooks}). Although resolution effects on CR-CD runs are shown to be relatively small (see \ses{3.1} and \ref{sec:4.1.3}), they still may affect the variant CR runs, and the X-ray bubble formation (see \se{4.2.1}) may still be sensitive to resolutions, which can impact the results of the X-ray emission features. \textit{(3). Lack of statistics.} In this work, we exclusively focus on one MW-like galaxy in FIRE simulations. To gain better statistics and to better match the observational surveys that span a wider range of halo masses, similar analysis needs to be done on a larger sample of simulated halos. Although we tested our results on several slightly different haloes (see Appendix \ref{apx:addtional}), we stress that halo-to-halo differences may be much more significant than what we found and should be further investigated in future studies. 
    \end{itemize}
    
    \item \textbf{Summary on viability of CR models:} Though the analysis we have done in this work has demonstrated variations between different models and provides insights on how they affect galaxy evolution, the CR models are still mostly unconstrained. The CR-CD run, though not having a clear physical motivation, still exhibits features in agreement with most observational constraints. As for the variant diffusivity CR runs, the SC model exhibits dramatic instability in galaxies of $M_{\rm halo}\sim 10^{12}\Msun$ and should be treated with caution in future work. One of the robust conclusions we can draw is that CRs can lower the temperature and affect the phase structure, resulting in higher UV ion values and lower thermal $L_{\rm X}$ values. Distinguishing between the thermal and the ICS contributions to the X-ray emission from the CGM would greatly help constrain the transport models.
    
    \item \textbf{Other FIRE CR runs for other host haloes (not presented in the main text): } Complementarily, we find no significant halo-to-halo variations on any of the aforementioned results, and the details are discussed in Appendix \ref{apx:addtional}. However, we only have a handful of additional zoom-in simulations and, therefore, cannot make a close match to the observed samples, preventing us from making stronger conclusions in this work.

\end{enumerate}

\subsection{Outlook}
\label{sec:5.2}
Given that the largest difference between CR models comes from the extended CGM's thermal X-ray emission features (for gas in $r\gsim 0.25 R_{\rm vir}$, see \fig{X_ray_Anderson}), we expect any better separation by galaxy types and more reliable background subtraction in observations will enable stronger constraints to CR transport models in the future. Better constraints on the radial profile of this emission will potentially help distinguish between thermal and ICS contributions to the extended emission, further helping to constrain the CR transport models. Although no strong conclusion was reached in the present work, we suspect that further useful constraints will come from the combination of full-spectrum and CR-ET models run in high resolution. Runs with such combinations already existed with FIRE-3 physics and were used for synthetic observation of the FIR-radio correlation \citep{ponnada2025hooks}, but we do not yet have the complete set of transport model variations explored here run with the same FIRE feedback physics. Additionally, the non-linear ``blowout'' events of the CR-SC runs should be investigated more in future studies, to unlock the exact relation between the timing of the ``blowouts'' and the parameters of the CR model, mass resolution, or other galaxy/numerical properties. Lastly, as briefly discussed in this paper, resolution effects can be potentially important in the cold cloud formation in CGM, and therefore affect many CGM properties in a way that might differ between different CR transport models, so a dedicated study on resolution effects may be fruitful in the future.

\section*{Acknowledgements}
We thank the anonymous referee for constructive comments that improved the quality of this manuscript. YSL thanks Caleb Choban, Cameron Trapp, Suoqing Ji, and TK Chan for technical help and useful discussions. YSL and DK were supported by the National Science Foundation (NSF) grant AST-2108314. Support for PFH \&\ SP was provided by NSF Research Grants 20009234, 2108318, NASA grant 80NSSC18K0562, and a Simons Investigator Award. CAFG was supported by NSF through grants AST-2108230 and AST-2307327; by NASA through grants 21-ATP21-0036 and 23-ATP23-0008; and by STScI through grant JWST-AR-03252.001-A. CBH was supported through NASA grants 80NSSC23K1515, HST-AR-15800, HST-AR-16633, and HST-GO-16703. The simulations presented here used computational resources granted by the Extreme Science and Engineering Discovery Environment (XSEDE), which is supported by National Science Foundation grant no. OCI-1053575, specifically allocation TG-AST120025 and resources provided by PRACNSF.1713353 supported by the NSF; Frontera allocations AST21010 and AST20016, supported by the NSF and TACC. The analyses were performed on the Triton Shared Computing Cluster (TSCC) at the San Diego Supercomputer Center (SDSC). The data used in this work were, in part, hosted on facilities supported by the Scientific Computing Core at the Flatiron Institute, a division of the Simons Foundation. This work was performed in part at the Aspen Center for Physics, which is supported by National Science Foundation grant PHY-2210452

\section*{Data Availability Statement}
The data supporting the plots within this article are available on reasonable request to the corresponding author. A public version of the GIZMO code is available at \hyperlink{http://www.tapir.caltech.edu/~ phopkins/Site/GIZMO.html}{http://www.tapir.caltech.edu/$\sim$ phopkins/Site/GIZMO.html}. Additional data including simulation snapshots, initial conditions, and derived data products are available at \hyperlink{http://fire.northwestern.edu}{http://fire.northwestern.edu}.


\bibliographystyle{mnras}
\bibliography{references.bib}



\appendix


\section{Star formation histories}
\label{sec:SF_histories}

\begin{figure*}
    \centering
    \includegraphics[trim={0.0cm 0.0cm 0.0cm 0.0cm}, clip, width =0.95 \textwidth]{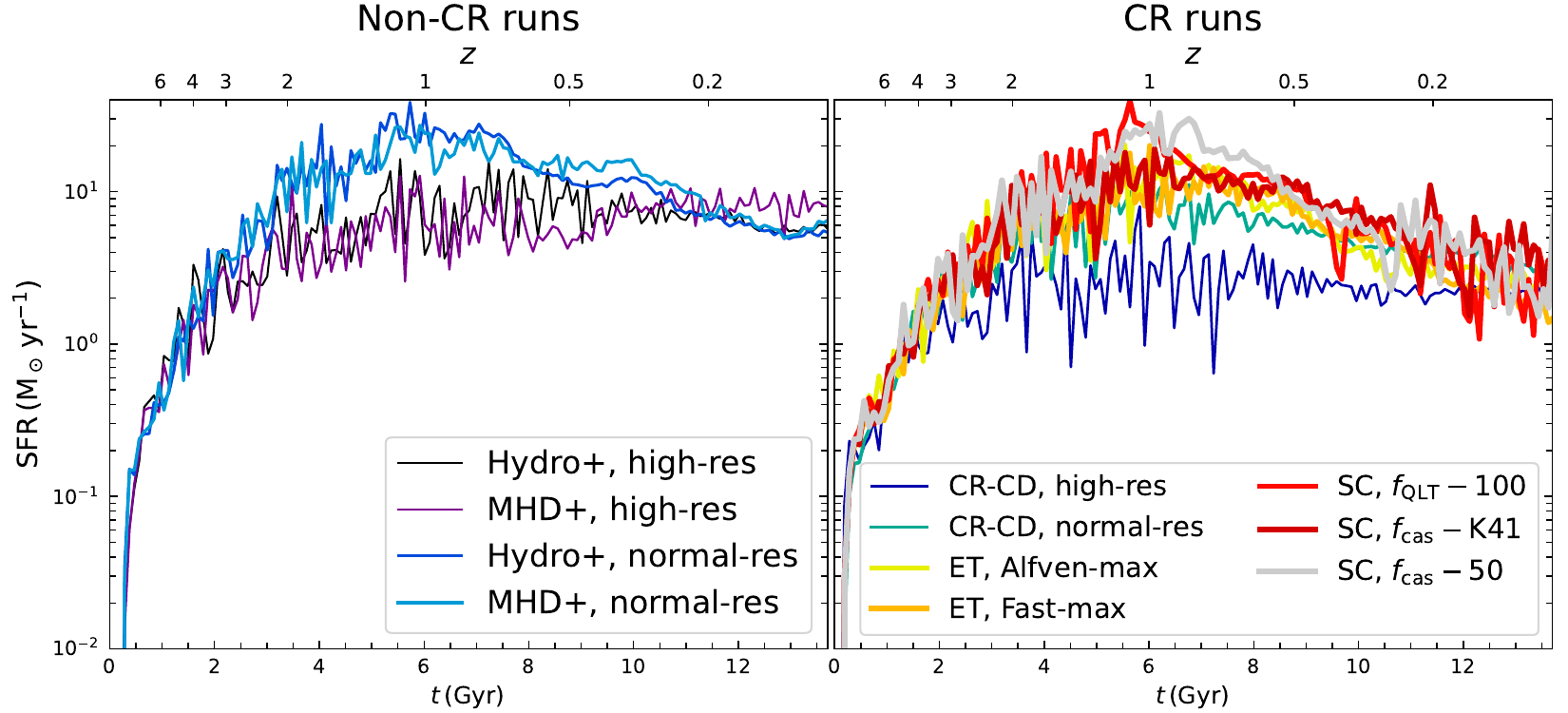}
    \caption{Archaeological star formation rates (SFRs) (using only present day stellar mass and neglecting stellar mass loss; also used for \fig{X_ray_Li}). All runs reach their peaked SFR at $z\sim 2-1$. At $z\sim 0$, all CR runs have similar SFRs while non-CR runs have overall higher SFRs.}
    \label{fig:SF_histories}
\end{figure*}

In \fig{SF_histories} we show the star formation histories of each of our simulation runs. The normal-res Hydro+ and MHD+ runs have high SFRs ($\gsim 20 \, \msun \yr^{-1}$) at $z\sim 1-2$, dropping to $\sim 5 \, \msun \yr^{-1}$ by $z=0$, resulting in high stellar masses at $z\sim 0$ as shown in \tab{sims}. SFRs are lower at early times in the high-resolution Hydro + and MHD + runs than in normal-resolution runs, but they end up being higher at late times ($z\lsim 0.2$), because there is more gas available for star formation (see \fig{proj_resolution_comparison}). At $z\sim 0$, all non-CR runs have overall higher SFRs than CR-CD runs, which indicates that CRs can largely suppress star formation at late time in \texttt{m12} simulations (also see discussion in \citet{hopkins2020but}). This is mainly due to the additional non-thermal supports from CRs that drive more outflows in the CGM and slow down the gas infall from the CGM \citep{hopkins2021cosmic, chan2022impact}, reducing the fuel for galactic star formation. Owing to a combination of differences in their star formation histories, and local and global influence of CRs on the SFRs, no significant differences are observed between runs with different CR transport models at $z\sim 0$.

\section{The ``Blowout'' of Galaxies with Self-confinement CR Models}
\label{apx:investigation}

\begin{figure*}
    \centering
    \includegraphics[trim={0.0cm 0.0cm 0.0cm 0.0cm}, clip, width =0.98 \textwidth]{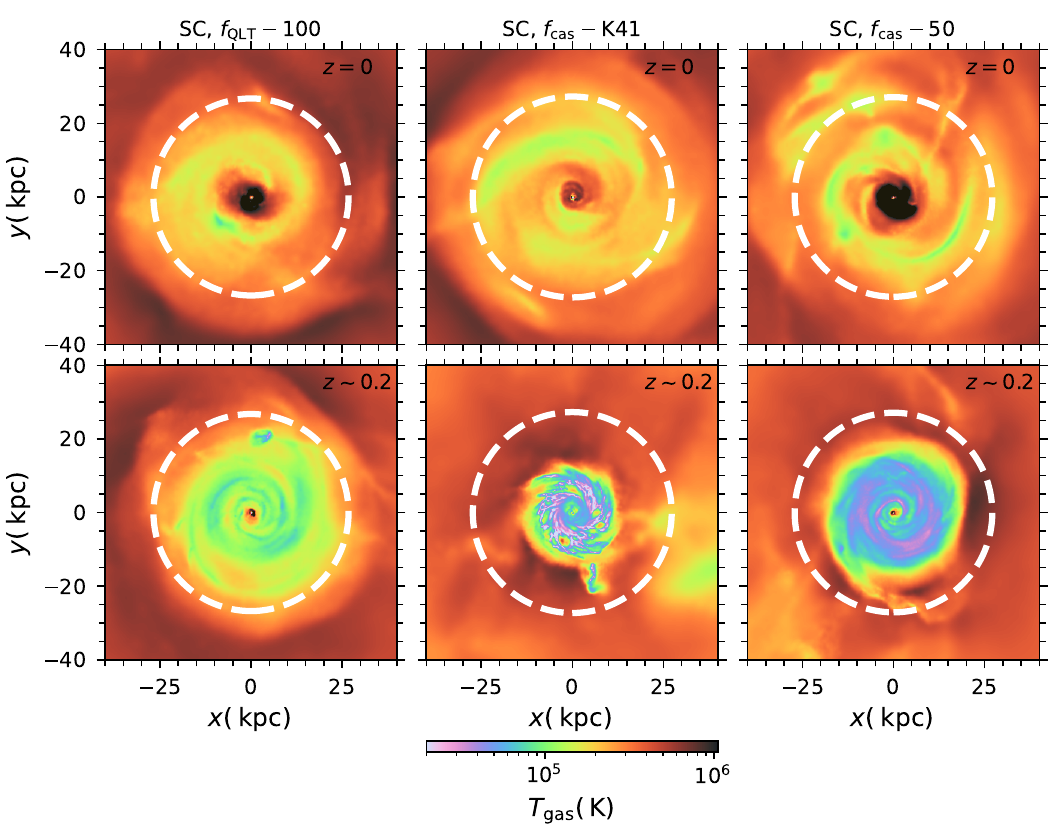}
    \caption{Density-weighted temperature projection maps for all SC runs, with the specific models labelled at the top. We present only the face-on view of each galaxy. We compare projection maps between $z=0$ and $z\sim 0.2$. Model-to-model variations are clearly seen, as the exact timing for the blowout to happen can depend on not only the parameters of CR transport, but also the numerical details and host galaxies.}
    \label{fig:Temp_proj_SC}
\end{figure*}

In \se{3.1} we mentioned the ``blowout''-like events driven by instability of CR diffusion in the SC models. Here, to aid our arguments in the main text, we show an earlier snapshot of the three SC runs to further illustrate how the ``blowout'' event occurs.

\Fig{Temp_proj_SC} shows the projection maps of the gas temperature of our CR-SC runs. The upper panels are snapshots at $z=0$ and the lower panels are snapshots at $z\sim 0.2$. All panels are the ``face-on'' views. Hot regions are remarkable in the centres of galaxies at $z=0$, whereas at $z\sim 0.2$ all three galaxies have more cold gas compared to $z=0$, and no significant central ``hot region''.  Among three simulations, $f_{\rm cas}-{\rm K41}$ appears to have a most ``intact'' disk at $z\sim 0.2$, potentially indicating that the ``blowout'' happened the latest for this run. At a similar time, the $f_{\rm cas}-50$ and $f_{\rm QLT}-100$ runs appear to be undergoing a ``blowout'' event, with a central ``hole'' filled with very hot gas that increases in size while clearing most of the cold disk gas by $z=0$. Further investigation is needed to understand the timing and galaxy mass/type dependence of this instability of our SC transport scheme. 

\section{Additional FIRE Galaxies with CRs}
\label{apx:addtional}

\begin{figure*}
    \centering
    \includegraphics[trim={0.0cm 0.0cm 0.0cm 0.0cm}, clip, width =0.98 \textwidth]{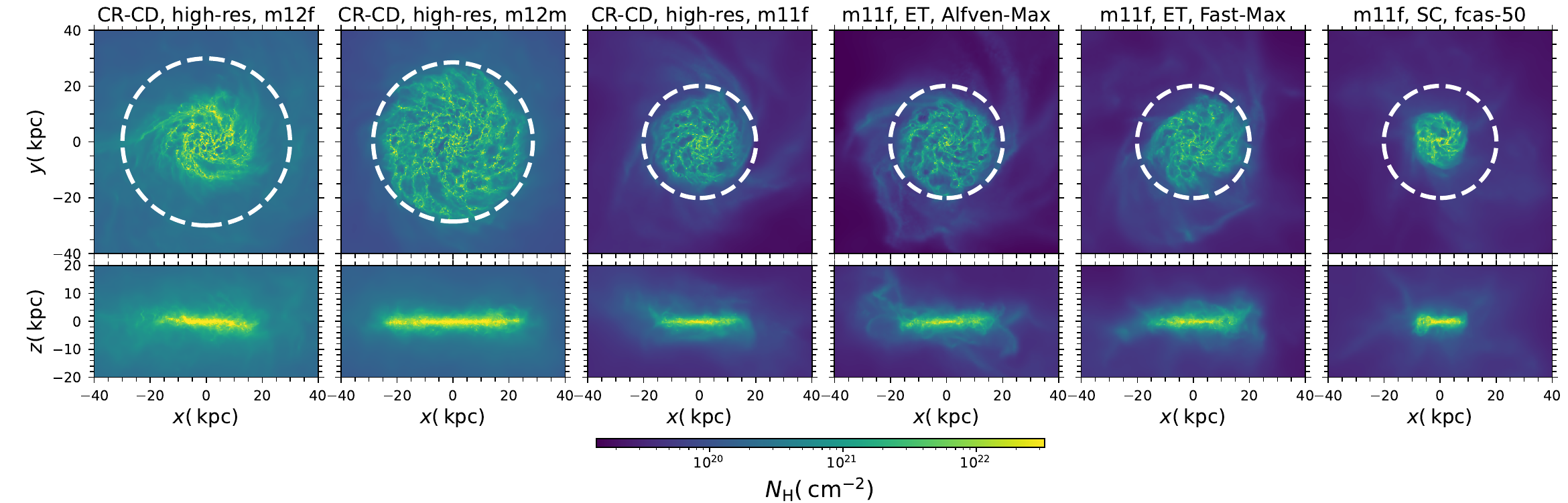}
    \caption{Projection maps of the hydrogen column density $N_{\rm H}$ for our additional runs. We label the each run with their corresponding galaxy and CR model at the top of each panel. The presentation of the plot is in the same manner as \fig{proj_variant_kappa_CR}. We notice that all galaxies have similar disk structure. \texttt{m12m} and \texttt{m12f} have more massive and larger disks, whereas \texttt{m11f} runs are less massive and smaller. Those are consistent with their halo masses.}
    \label{fig:Additional_projs}
\end{figure*}

\begin{figure*}
    \centering
    \includegraphics[trim={0.0cm 0.0cm 0.0cm 0.0cm}, clip, width =0.98 \textwidth]{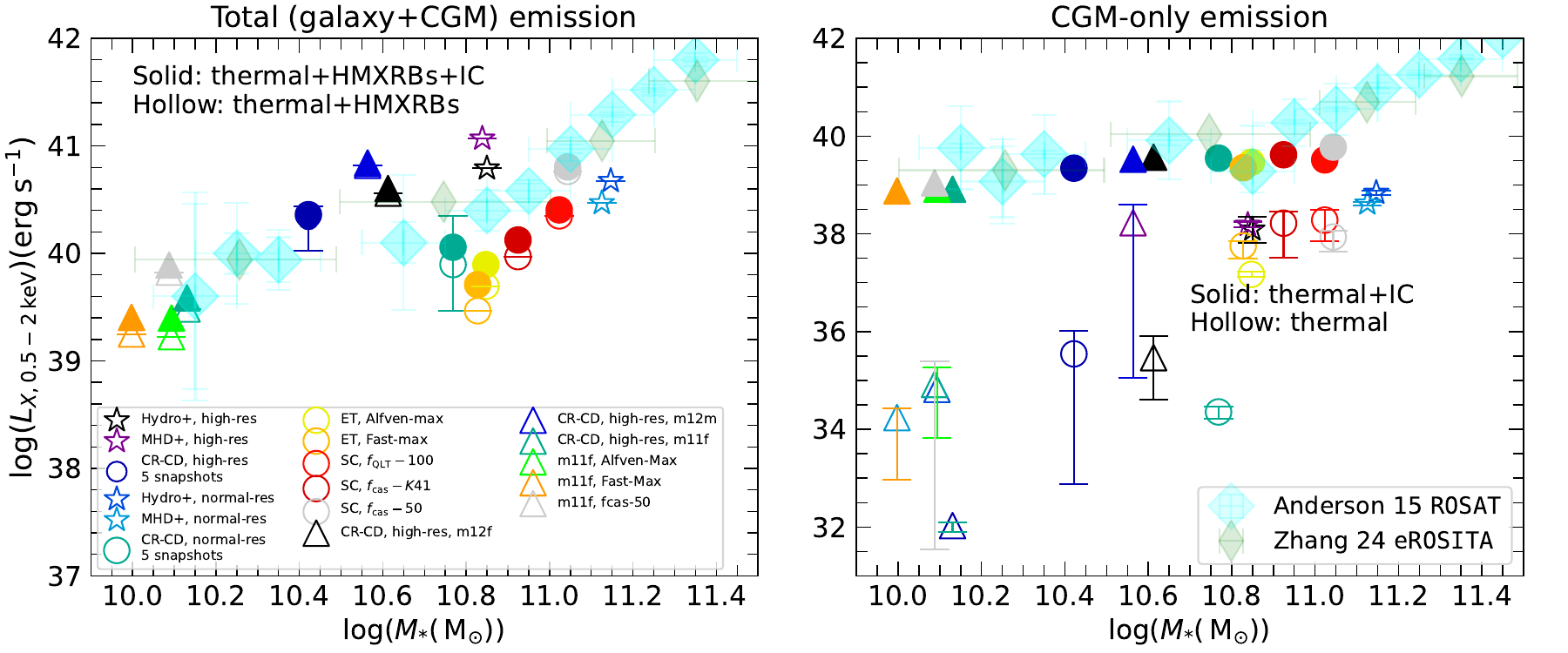}
    \includegraphics[trim={0.0cm 0.0cm 0.0cm 0.0cm}, clip, width =0.46 \textwidth]{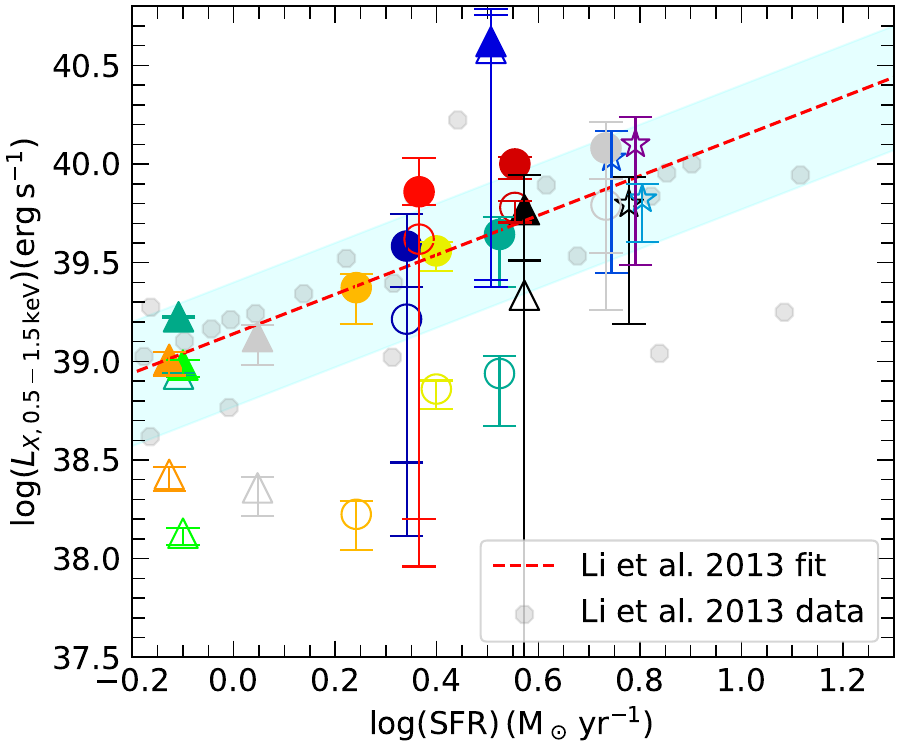}
    \caption{Results from Figures  \ref{fig:X_ray_Anderson} and \ref{fig:X_ray_Li} (the top and bottom panels, respectively) with the additional runs included. Except for the new data points (the triangular shaped points), everything else is exactly the same as the corresponding plots in the main text. See the legend for the corresponding galaxies and models.}
    \label{fig:Extra_X_ray}
\end{figure*}

In the main text, all runs have a host galaxy of \texttt{m12i}. In this section, we supplement some of the figures in the main text with additional simulations. We do not have a complete set of CR transport physics available, but we can still use them to test the robustness of our main results. A total of six (6) additional runs are presented and are listed below.

\begin{enumerate}
    \item \texttt{m12m}, high-resolution, CR-CD;
    \item \texttt{m12f}, high-resolution, CR-CD;
    \item \texttt{m11f}, high-resolution, CR-CD;
    \item \texttt{m11f}, normal-resolution, ET, Alfvén-Max;
    \item \texttt{m11f}, normal-resolution, ET, Fast-Max;
    \item \texttt{m11f}, normal-resolution, SC, $f_{\rm cas}-50$.
\end{enumerate}

In \fig{Additional_projs} we show the projected face-on and edge-on views of the hydrogen column density $N_{\rm H}$ of the additional runs. Unsurprisingly, both the \texttt{m12f} and \texttt{m12m} CR-CD runs look similar to the \texttt{m12i} CR-CD run. The \texttt{m11f} CR-CD run has a slightly smaller disk because the galaxy is less massive (an intermediate-mass dwarf). The disk sizes in the two ET runs for \texttt{m11f} are similar to the CR-CD run, whereas in the \texttt{m12i} case the disks are obviously smaller in the ET runs. This suggests that some of the disk size differences we see are subject to simple run-to-run variations and are not generally applicable (see  \se{3.1}). Another result not following the \texttt{m12i} patterns is that the \texttt{m11f} SC run has some visible disk structure even at $z=0$, unlike the \texttt{m12i} SC runs, which have the central gas region completely blown away by then. This aids our argument in \se{3.1} and Appendix \ref{apx:investigation} that these events are highly non-linear and the exact timing for the ``blowout'' to occur potentially depends on the disk formation time.

In \fig{Extra_X_ray} we show our three main X-ray results, with the triangular shapes representing the additional runs not included in the main text. The details of the figure, including the observational data, remain the same as in Figures \ref{fig:X_ray_Anderson} and \ref{fig:X_ray_Li}. Similar to the results in our main text, the ICS dominates the thermal emission in the CGM, but not in the total emission that includes the central galaxy. Since \texttt{m12m} and \texttt{m12f} galaxies are similar to our main \texttt{m12i} galaxy, those two runs are close to the \texttt{m12i} data points. The \texttt{m11f} galaxies, on the other hand, are less massive and therefore have lower SFRs. However, the difference between the \texttt{m11f} CR-CD and its CR variants is not as large. These verify the robustness of our results in terms of X-ray emissions.

Note that we did not present results of thermal profiles and the ion absorption columns for these additional galaxies. The latter has already appeared in \citet{ji2020properties} and no clear halo-to-halo difference was spotted. As for the thermal profiles, the density and pressure are very similar between our additional galaxies and the galaxies presented in the main text, and the temperature profiles follow similar trends i.e. for the same galaxy, the ordering of temperature profiles from highest to lowest temperature is non-CR > ET$\approx$SC > CR-CD.

\bsp	
\label{lastpage}
\end{document}